\documentclass[sigplan,10pt,
screen,
]{acmart} 

\usepackage{xurl}

\AtBeginDocument{%
 }

\PassOptionsToPackage{hyphens}{url}
\usepackage{cleveref}
\usepackage{listings}
\usepackage{enumitem}
\usepackage{microtype}
\usepackage{breakurl}

\usepackage[colorinlistoftodos]{todonotes}

\definecolor{keywordcolor}{rgb}{0.7, 0.1, 0.1}   
\definecolor{commentcolor}{rgb}{0.4, 0.4, 0.4}   
\definecolor{errorcolor}{rgb}{0.9, 0.0, 0.0}     
\definecolor{symbolcolor}{rgb}{0.4, 0.4, 0.4}    
\definecolor{sortcolor}{rgb}{0.1, 0.5, 0.1}      
\definecolor{stringcolor}{rgb}{0.31, 0.71, 0.14}      

\usepackage{xspace}
\newcommand{\mathlib}{Mathlib\xspace}

\newcommand{\CC}{\mathbf{C}}
\newcommand{\FF}{\mathbf{F}}
\newcommand{\NN}{\mathbf{N}}
\newcommand{\QQ}{\mathbf{Q}}
\newcommand{\ZZ}{\mathbf{Z}}
\newcommand{\OO}{\mathcal{O}\xspace}

\newcommand{\OK}{\mathcal{O}_K}

\DeclareMathOperator{\Disc}{Disc}

\DeclareMathOperator{\Res}{Res}

\DeclareMathOperator{\Fr}{Frob}

\usepackage{MnSymbol}

\copyrightyear{2025}
\acmYear{2025}
\setcopyright{cc} 
\setcctype{by-nc-sa}
\acmConference[CPP '25]{Proceedings of the 14th ACM SIGPLAN International Conference on Certified Programs and Proofs}{January 20--21, 2025}{Denver, CO, USA}
\acmBooktitle{Proceedings of the 14th ACM SIGPLAN International Conference on Certified Programs and Proofs (CPP '25), January 20--21, 2025, Denver, CO, USA}
\acmDOI{10.1145/3703595.3705874}
\acmISBN{979-8-4007-1347-7/25/01}

\begin{document}

\title{Certifying Rings of Integers in Number Fields}


\author{Anne Baanen}
\authornote{Authors listed in alphabetical order}
\email{anne@anne.mx}
\orcid{0000-0001-8497-3683}
\affiliation{%
	\institution{Vrije Universiteit Amsterdam}
	\city{Amsterdam}
	\country{Netherlands}
}
\affiliation{%
	\institution{Lean FRO, USA}
	\country{}
}

\author{Alain Chavarri Villarello}
\orcid{0009-0006-5203-4781}
\email{a.chavarri.villarello@vu.nl}
\affiliation{%
	\institution{Vrije Universiteit Amsterdam}
	\city{Amsterdam}
	\country{Netherlands}
}

\author{Sander R. Dahmen}
\orcid{0000-0002-0014-0789}
\email{s.r.dahmen@vu.nl}
\affiliation{%
	\institution{Vrije Universiteit Amsterdam}
	\city{Amsterdam}
	\country{Netherlands}
}


\begin{abstract}
Number fields and their rings of integers, which generalize the rational numbers and the integers, are foundational objects in number theory. 
There are several computer algebra systems and databases concerned with the computational aspects of these.
In particular, computing the ring of integers of a given number field is one of the main tasks of computational algebraic number theory. 
In this paper, we describe a formalization in Lean 4 for certifying such computations. 
In order to accomplish this, we developed several data types amenable to computation.
Moreover, many other underlying mathematical concepts and results had to be formalized, most of which are also of independent interest.
These include resultants and discriminants, as well as methods for proving irreducibility of univariate polynomials over finite fields and over the rational numbers.
To illustrate the feasibility of our strategy, we formally verified entries from the \emph{Number fields} section of the \emph{L-functions and modular forms database} (LMFDB).
These concern, for several number fields, the explicitly given \emph{integral basis} of the ring of integers and the \emph{discriminant}.
To accomplish this, we wrote SageMath code that computes the corresponding certificates and outputs a Lean proof of the statement to be verified. 
\end{abstract}



\maketitle

\section{Introduction}\label{sec:intro}

\looseness=-1
There are many fundamental concepts in mathematics that are, in principle, amenable to computation.
Focusing on number theory, and more particularly algebraic number theory 
(i.e. the theory of algebraic numbers), 
such concepts include the \emph{rings of integers} of a \emph{number field}.
Number fields generalize the field of rational numbers $\QQ$ (in the sense that they are finite degree field extensions thereof) and each contains a ring of integers, which can be seen as generalizing how the (ordinary) integers $\ZZ$ are contained in the number field $\QQ$; see Section~\ref{sec:prelim} for more details.
Rings of integers are thereby key to the arithmetic properties of their number fields,
and both are essential to number theory from the 19th century to the modern day.
Their basic definitions and properties (amongst other concepts) were formalized in~\cite{class-group-jar}, on which we build.
From a computational direction, number fields of degree $2$ and their rings of integers were considered in~\cite{diophantine-toolkit-cpp}.
This paper can be seen as a far reaching generalization of such formally verified computations.
In order to work with rings of integers inside a proof assistant ---Lean 4 in our case--- we opted for a \emph{certification} approach.
Given a ring of integers, concretely represented by giving an \emph{integral basis} (i.e. a $\ZZ$-basis) for it, as e.g. computed by an external Computer Algebra System (CAS) or extracted from some database, we let SageMath~\cite{SageMath} compute a certificate which is checked in Lean to certify the correctness of the ring of integers. In fact, our SageMath code outputs a complete proof which Lean can readily check.

The isomorphism class of a number field can be represented explicitly, for example by a polynomial with integer coefficients and leading coefficient $1$ that is irreducible over $\ZZ$ (and hence $\QQ$).
This is also a basic way to define number fields in Computer Algebra Systems such as \mbox{PARI}/GP~\cite{PARI2}, SageMath~\cite{SageMath}, and Magma~\cite{Magma} (perhaps relaxing to rational coefficients).
Given the prime factorization of the discriminant of the defining polynomial, there exist efficient algorithms to determine the ring of integers, e.g.~\cite[Algorithm 6.1.8]{Cohen}, \cite[Chapter III]{Ford},  and implementations are available in various Computer Algebra Systems, including those mentioned before.
However, these algorithms are quite involved, which is one reason why we opted for a \emph{certification} approach for rings of integers.
This is described in Section~\ref{sec:rings-of-integers}, the technical heart of this paper.

To show the feasibility of our approach, we formally verified several entries in the well known L-functions and modular forms database (LMFDB)~\cite{lmfdb}.
For various number fields, we verified in Lean the ring of integers, as given by an explicit \emph{integral basis}, as well as the \emph{discriminant} (an integer valued invariant); see Section~\ref{sec:lmfdb}.

In Section~\ref{sec:computation}, we provide an overview of the different notions of Lean computation used.
In order for everything to function, we needed to formalize much underlying theory and computational methods, most of which are interesting independently.
Notable topics include irreducibility of polynomials over $\QQ$ and finite fields (see Section~\ref{sec:irreducibility}), as well as resultants and discriminants (see Section~\ref{sec:resultants-discriminants}).

We conclude the paper with a brief discussion (in Section~\ref{sec:discussion}), including related and future work.
Full source code of our formalization and SageMath scripts are available.\footnote{\url{https://github.com/alainchmt/RingOfIntegersProject}} 

\section{Preliminaries}\label{sec:prelim}

In this paper, we assume some familiarity with basic ring and field theory, as can be found e.g. in the undergraduate textbooks~\cite{Dummit-and-Foote,lang2005undergraduate}.
For sake of self-containedness, we will discuss in this section some basics concerning number fields and their rings of integers; the following paragraph essentially follows the exposition (with very little modification) from~\cite[Section 2]{diophantine-toolkit-cpp}.

A number field $K$ is a finite extension of the field $\QQ$.
It is a finite dimensional vector space over $\QQ$, and its dimension is called the degree of $K$ (over $\QQ$).
Examples of number fields are $\QQ$ itself (of degree $1$), $\QQ(\sqrt{3}) = \{a+b\sqrt{3} : a,b \in \QQ\}$ and $\QQ(\sqrt{-3}) = \{a+b\sqrt{-3} : a,b \in \QQ\}$ (both of degree 2), and
$\QQ(\alpha)=\{a+b \alpha+c \alpha^2 : a,b,c \in \QQ\}$ where $\alpha$ is any (complex) root of  the polynomial $x^3-3 x-10$.
This latter example generalizes as follows.
For any polynomial of degree $n$ that is irreducible over $\QQ$, adjoining any one of its (complex) roots $\beta$ to $\QQ$ will yield the degree $n$ number field
$\QQ(\beta) = \{a_0+a_1 \beta + \ldots + a_{n-1} \beta^{n-1} : a_0, a_1, \ldots, a_{n-1} \in \QQ\}$ (the $n$ different choices of $\beta$ will all yield isomorphic fields).
Conversely, any number field of degree $n$ will be isomorphic to such a field $\QQ(\beta)$.
From an algebraic and arithmetic perspective, the (rational) integers $\ZZ$ constitute a particularly \lq nice\rq\ subring of its field of fractions, the rational numbers $\QQ$.
Upon generalizing from $\QQ$ to an arbitrary number field $K$, the analogue of $\ZZ$ is the ring of integers of $K$, denoted $\OK$.
It is defined as the integral closure of $\ZZ$ in $K$:
\[ \OK := \{x \in K :  \exists f \in \ZZ[x] \text{ monic}  : f(x)=0\}\]
where we recall that a (univariate) polynomial is called \emph{monic} if its leading coefficient is equal to 1.
The fact that $\OK$ is indeed a ring follows for instance from general properties of integral closures (\cite[Section I.2]{Neukirch}).

The ring of integers for the four explicit examples of number fields above, are
$\OO_{\QQ}=\ZZ$ (indeed), $\OO_{\QQ(\sqrt{3})}=\ZZ[\sqrt{3}]=\{a+b\sqrt{3}: a,b \in \ZZ\}$,
$\OO_{\QQ(\sqrt{-3})}=\ZZ[(1+\sqrt{-3})/2]=\{a+b(1+\sqrt{-3})/2: a,b \in \ZZ\}$,
and $\OO_{\QQ(\alpha)}=\{a+b\alpha+c(\alpha-\alpha^2)/2 : a,b,c \in \ZZ\}$.
For the third example, note that indeed $(1+\sqrt{-3})/2 \in \OO_{\QQ(\sqrt{-3})}$ as it is a root of the monic polynomial $x^2-x+1$.

The last (cubic) example shows that the ring of integers can become complicated very quickly.
In this case $\OO_{\QQ(\alpha)}$ \emph{cannot} be written in the form $\ZZ[\gamma]$ for any element $\gamma \in \OO_{\QQ(\alpha)}$, which is phrased as $\OO_{\QQ(\alpha)}$ being not \emph{monogenic}.
Note that any number field (say of degree $n$) is monogenic, as it can be generated by a single element $\beta$ over $\QQ$, which yields $\{1,\beta, \beta^2,\ldots, \beta^{n-1}\}$ as a basis for the corresponding $\QQ$-vector space, called a \emph{power basis}.
Observe that the ring of integers $\OK$ of a number field $K$ is a $\ZZ$-module. Recall that in general an $R$-module is just a vector space if $R$ is a (skew) field, but the same standard defining axioms make sense in the more general case that the \emph{scalars} $R$ are only assumed to form a ring.
While the $\ZZ$-module $\OK$ might not have a power basis (as it may not be monogenic), it always has some basis, called a $\ZZ$-basis or \emph{integral basis}.
E.g. in the cubic example, a $\ZZ$-basis is given by $\{1,\alpha, (\alpha-\alpha^2)/2\}$, as any element of $\OO_{\QQ(\alpha)}$ can be expressed uniquely as a $\ZZ$-linear combination of these three basis elements.
Finally, if we have a subring $\OO$ of a number field $K$, which as a $\ZZ$-module has a basis that also forms a $\QQ$-basis for $K$, then $\OO$ is actually contained in $\OK$.

\section{Forms of Computation in Lean} \label{sec:computation}
This section provides an overview of the different notions of computation that we had to deal with.

Lean has a built-in notion of computation through \emph{reduction}:
this is a primitive relation between two terms reflecting the computational content of the Calculus of Inductive Constructions~\cite{lean-tt}.
Reduction is invoked as part of definitional equality checking,
and allows Lean to verify equalities such as $2 + 2 = 1 + 3$ through computing that either side evaluates to $4$.
Definitionally equal terms are indistinguishable for the type theory,
and therefore a proof of $2 + 2 = 1 + 3$ can be given by the reflexivity principle of equality, \lstinline{rfl}.
In Lean 4, reduction also applies to~
literals of primitive types: natural numbers and strings.
Reduction is implemented as part of the trusted codebase in the Lean kernel.

Lean as a programming language additionally possesses a notion of computation for definitions
which can be compiled into executable code and \emph{evaluated}.
Evaluation is somewhat orthogonal to reduction in that it is not part of the logic of Lean,
and not all definitions that can be evaluated will reduce, and vice versa.
In particular, Lean includes --and \mathlib makes consistent use of-- the axiom of choice,
resulting in \emph{noncomputable} definitions that cannot be evaluated.
The user can additionally override with arbitrary code the computational meaning of a definition,
using the \lstinline{@[implemented_by]} attribute.

Evaluation is the notion of computation powering the tactic framework.
Broadly speaking, tactics are evaluated Lean code that operate directly on the term level.
Tactics can also be said to perform computation, by taking in a term \lstinline{t} and producing a new term \lstinline{t'} alongside a proof term of type \lstinline{t = t'}.
The \lstinline{norm_num} tactic for computing with numerical expressions is an example of this design.
Often, a tactic works by \emph{reflecting} a given term into a data structure internal to the tactic,
using computation on internal data to construct a proof term.
Since tactics can choose the internal representation of the data structures,
they reimbue noncomputable terms with computational meaning.
An example is the \lstinline{ring} tactic of \mathlib which computes on polynomial expressions
although the ring structure of the type \lstinline{Polynomial} itself is marked \lstinline{noncomputable}.

If a tactic returns \lstinline{rfl} as the proof term for the equality \lstinline{t = t'},
verifying the definitional equality of the terms \lstinline{t} and \lstinline{t'} can invoke reduction:
this principle of \emph{computational reflection}~\cite{reflection-tactics}
saves the need to go through the computation twice, once on the tactic-internal data structure and once on the constructed proof term.
The tactic \lstinline{decide} is an example of a reflective tactic in Lean,
proving a given goal \lstinline{P} by finding a corresponding decision procedure \lstinline{d} such that \lstinline{d = true} if and only if \lstinline{P},
and then verifying that \lstinline{d = true} holds.
This tactic has two variants: \lstinline{decide} first computes whether \lstinline{d = true} in the elaborator, before submitting the same computation to the kernel,
while \lstinline|decide!| directly invokes kernel reduction and thus can unfold every definition.

Efficiency is an important consideration in the use of computation in Lean.
In the highest generality, evaluation is faster than reduction since evaluation does not rely on the abstraction of terms
that must be constructed and matched for each reduction step.
The higher efficiency of evaluation comes at the cost of being separated from the logic: proof checking can only invoke reduction.
Lean provides an escape hatch in the form of the \lstinline{native_decide} tactic:
unlike \lstinline{decide}, this allows a proof whenever a decision procedure \emph{evaluates} to \lstinline{true}.
Use of \lstinline{native_decide} not only means that the entire evaluation mechanism has to be added to the trusted codebase,
in addition the correctness of all \lstinline{@[implemented_by]} attributes must be trusted.
Use of \lstinline{native_decide} is therefore a tradeoff between efficiency and certainty.
See Section~\ref{sec:lmfdb} for some performance measurements in our project.

\section{Irreducibility}\label{sec:irreducibility}
\looseness=-1
For a primitive polynomial $f$ in $\ZZ[X]$ -- a polynomial whose coefficients have greatest common divisor equal to $1$-- proving that $\QQ[X]/ \langle f \rangle$ is a number field is equivalent to proving that $f$ is irreducible over $\ZZ[X]$. Thus, generating an irreducibility proof for polynomials over the integers in Lean is crucial for our purposes. 

\subsection{Polynomials as Lists}~\label{sec:asLists}
\looseness=-1
In \mathlib, the underlying representation of polynomials is implemented in such a way that polynomial arithmetic is noncomputable. Therefore, we work with an alternative representation of polynomials as lists. 

For a semiring $R$ with decidable equality, we define a map \lstinline|Polynomial.ofList : List R → R[X]| that converts a list of coefficients $[a_0, \ldots, a_n]$ into the polynomial $a_0 + a_1 X + \ldots + a_n X^n$. We then define computable operations on \lstinline|List R| that correspond to operations on polynomials, such as addition and multiplication. This translation of polynomial arithmetic into list arithmetic will be our general strategy for dealing with polynomial computations. For example, to prove that \lstinline|ofList|~$ l_1$ \lstinline|* ofList|~$ l_2$ \lstinline| = ofList| $l_3$, we show: 
\begin{lstlisting}
((*$l_2$*) * (*$l_2$*)).dropTrailingZeros = (*$l_3$*).dropTrailingZeros
\end{lstlisting}
This equality can be proven using the \lstinline|rfl| and \lstinline|decide| tactics. Here, \lstinline|dropTrailingZeros| removes the trailing zeros of a list. 

We note that this approach to polynomial computation relies on the input polynomials being provided in the list-based format,
as the standard way to define polynomials in \mathlib is noncomputable.
Alternatively, one could consider using the \mathlib representation of polynomials along with tactics such as \lstinline|ring| \cite{baanenring} (which is partly based on the Coq tactic \cite{ring}) and \lstinline|norm_num| extensions like \lstinline|reduce_mod_char| to prove equalities involving additions and multiplications. However, this method proved too slow for our purposes, which led us to adopt the list-based approach instead.  

In addition to \lstinline|Polynomial.ofList| and the arithmetic operations on lists, we defined \lstinline|ComputablePolynomial R| as the subtype of lists $l$ over \lstinline|R| such that \lstinline|l = l.dropTrailingZeros|, and proved that it is isomorphic to \mathlib polynomials. The algebraic structure of \lstinline|ComputablePolynomial R|, namely, that it is a ring when \lstinline|R| is a ring, is pulled from the corresponding instance in \mathlib polynomials. 
This representation is similar to the one in Coq's Mathematical Components library \cite{Assia}, which uses lists with a non-zero last entry to define polynomials, though our approach does not assume that $R$ is nontrivial. 

Having multiple encodings for the same mathematical object, one better suited for proving theoretical properties and another for computation, has been explored in previous formalization efforts of algebraic algorithms \cite{maxime}. While we do not employ a proof transfer framework such as Trocq \cite{trocq}, this could potentially be explored to automate translations between equivalent data structures.

\subsection{Over Finite Fields}\label{finitefields}
\looseness=-1
In this section, we describe a way to prove irreducibility of polynomials over a finite field, which will aid us in proving irreducibility in $\ZZ[X]$. We then present a certificate for this irreducibility and its implementation in Lean. 

A finite field is a field with finite cardinality. This cardinality is necessarily a prime power and two finite fields of equal cardinality are always isomorphic. Thus, we may speak of \textit{the} finite field of cardinality $q$, which we denote by $\FF_q$. Remarkably, the irreducible factors of $X^{q^n} - X$ in $\FF_q[X]$ are precisely the irreducible polynomials in $\FF_q[X]$ that have degree dividing $n$. 
This gives rise to an irreducibility test due to Rabin \cite{Rabin}, which forms the basis to our formal irreducibility proofs. 
\begin{theorem}[Rabin's test]  \label{rabin}
	Let $f$ be a polynomial over $\FF_q$ of degree $n > 0$. 
	 Then $f$ is irreducible if and only if: 
	\begin{enumerate}
		\item $f$ divides $X^{q^n} - X$
		\item $\gcd(f, X^{q^{n/t} }- X) = 1$ for every prime $t$ dividing $n$. 
	\end{enumerate}
\end{theorem}

In Lean, to talk about a finite field, we follow the practice in \mathlib and endow \lstinline|(F : Type*)| with instances \lstinline|[Field F] [Fintype F]|. The instance \lstinline|[Fintype F]| carries data: a finite set containing all elements of \lstinline|F|. We use it rather than \lstinline|[Finite F]|, which merely asserts that \lstinline|F| is in bijection with \lstinline|Fin n|, the canonical type with \lstinline|n| elements, for some natural number \lstinline|n|. 
The previous statement --which we formalized-- reads: 
\begin{lstlisting}
theorem irreducible_iff_dvd_X_pow_sub_X' (f : F[X])
    (hd : 0 < f.natDegree) : Irreducible f ↔
  f ∣ (X ^ ((Fintype.card F) ^ f.natDegree) - X) ∧
  ∀ (p : ℕ), Nat.Prime p → p ∣ f.natDegree →
  IsCoprime f (X ^ (Fintype.card F ^ (f.natDegree / p)) - X)
\end{lstlisting}
The term \lstinline|Fintype.card F| is the cardinality of the finite set of all elements of \lstinline|F| coming from the \lstinline|[Fintype F]| instance. 

Currently, automatic computations in finite fields can only be carried out in Lean for fields of prime order (our work on irreducible polynomials can contribute to extending this to higher degree extensions), therefore we focus on the case $q = p$, with $p$ a prime number. In Lean, the integers modulo $p$ are represented as \mbox{\lstinline|ZMod p|}, which has decidable equality, allowing us to solve identity goals simply with the \lstinline|decide| tactic. Furthermore, if we have the instance \mbox{\lstinline|Fact (Nat.Prime p)|} available, then Lean infers that \lstinline|ZMod p| is a field.

Rabin's test starts by proving $f \mid X^{p^n} - X$, where $n$ is the degree of the polynomial $f$ in $\FF_{p}[X]$. The obvious way to do this is to provide the factor $s$ such that  $X^{p^n} - X = f * s$ and verify the equality. There is, however, a problem. In general, $s$ will not be sparse and it will be of huge degree equal to $p^n - n$. This makes storing it and verifying the corresponding identity infeasible. The standard approach taken in computer algebra~\cite{CAS} is instead to show that $X ^ {p ^ n} \equiv X \pmod f $ using a  \textit{square-and-multiply} algorithm, reducing modulo $f$ at each step by performing polynomial division with remainder. 
We currently do not have an effective implementation of polynomial division with remainder in Lean. However, with our implementation of polynomials as lists, we can quickly verify multiplications. This suggest a certificate for $f \mid X^{p^n} - X$ based on the \textit{square-and-multiply} algorithm. 

This well-known algorithm performs fast exponentiation on any monoid. When applied to the ring $\FF_p[X]/\langle f \rangle$, the underlying proposition looks like this: 

\begin{proposition}  \label{sqandmul}
	Let $f$ and $g$ in $\FF_p[X]$ and let $m$ be a non-negative integer. Let $(\beta_i)_{i =0}^{s}$ be the binary digits of $m$ so that 
	$m = \sum_{i = 0} ^ s \beta_i 2 ^ i$. Set $y_s \equiv g ^ {\beta_s} \pmod f$ and $y_i \equiv  y_{i + 1} ^ 2 * g ^ {\beta_i} \pmod f$ for $i = s-1, \ldots, 0$.  Then $y_0 \equiv g ^ m \pmod f$. 
\end{proposition}

Computing $g^m \pmod f$ this way requires significantly fewer operations than repeated multiplication. The polynomials $y_i$ can be chosen with degrees less than $n$. Note that, when $a$ and $b$ do not have very large degree, we \emph{can} certify $a \equiv b \pmod f$ in the obvious way. 

The second step in Rabin's test asks us to prove $\gcd(f,$ $X^{p^m} - X) = 1$ 
for divisors $m$ of $n$. This also involves polynomials of large degree. However, by certifying $X^{p ^ m} \equiv h_m \pmod f$ for $1 \leq m \leq n$, with $h_m$ of degree less than $n$, we can instead prove  $\gcd (f , h_m - X) =1$. 
This can be done by providing polynomials $a_m$ and $b_m$ --computed using the extended Euclidean algorithm-- such that $a_m * f + b_m * (h_m - X) = 1$. 

Putting it all together, we get a certificate which can be verified using only polynomial addition and multiplication. 

\paragraph{Certificate}
Let $f$ in $\FF_p[X]$ of degree $n > 0$, let $(\beta_i)_{i = 0}^{s}$ be the binary digits of $p$, so $p = \sum_{i = 0}^{s} \beta_i 2^ i$. A certificate for the irreducibility of $f$ consists of the following data: 
\begin{itemize}
	\item An $(n + 1)$-tuple $(h_0, \ldots, h_n)$ over $\FF_p[X]$.
	\item An $n \times s $ matrix $(g_{ij})$ over $\FF_p[X]$.
	\item An $n \times (s + 1)$ matrix $(h'_{ij})$ over $\FF_p[X]$.
	\item $n$-tuples $(a_0, \ldots, a_{n-1})$ and $(b_0, \ldots, b_{n-1})$ over $\FF_p[X]$.
\end{itemize}
The verification proceeds by checking the following statements:
\begin{enumerate}[label=(\roman*)]
	\item  For all $0 \leq i < n$, $h'_{is} =  h_i ^ {\beta_s}$ and $h'_{i0} = h_{i + 1}$. 
	\item  For all $0 \leq i < n$ and  $ 0 \leq j < s$,\\
	\hspace*{1em} $f * g_{ij} = {h'_{i(j+1)}}^2* h_i ^ {\beta_j} - h'_{ij}$.
	\item $h_0 = X$ and $h_n = X$. 
	\item  	$a_{n/t} * f + b_{n/t} * (h_{n/t} - X) = 1$ for every prime $t \mid n$.
\end{enumerate}
The first two statements in the verification imply that $h_i ^ p \equiv h_{i+1} \pmod f$ for all $0 \leq i < n$. Together with $h_0 = X$, we get that $X^{p^i} \equiv h_i \pmod f$ for every $0 \leq i \leq n$. Since $h_n = X$, the first part of Rabin's test is shown. The fourth statement proves the second part of the test. Note that in this last statement, only some entries $a_i$ and $b_i$, with $i$ dividing $n$, are used. Consequently, the rest of the entries in these $n$-tuples, which are included solely to simplify the indexing, can be set to zero. 

For our formalization, we use the following strategy repeatedly: given a certification scheme, we define a \lstinline|structure| in Lean where the fields include both the certifying data and the proofs of the verification statements. Often, we include more fields than what the informal description may suggest since some implicit assumptions require formal proof. 

For the above irreducibility certificate, we use our list-based approach to handle polynomial arithmetic. We bundle the certifying data and proofs of the verification statements into the structure: 
\begin{lstlisting}
structure CertificateIrreducibleZModOfList' (p n t s : ℕ) [Fact (Nat.Prime p)] [NeZero n] (L : List (ZMod p))
\end{lstlisting}

It is parametrized over $p$, $n$ and $s$ as before. Here, \lstinline|L| is the list of coefficients of the polynomial $f$. The parameter $t$ allows flexibility in choosing a base for the \textit{square-and-multiply}-type approach. We have described the method using the typical $t = 2$. However, choosing $t = p$ might be advantageous in some cases because computing $p$-powers in $\FF_p[X]$ is easy, although the $g_{ij}$ will have larger degree. 

As part of the fields in this structure, we include the factorization data for the degree $n$, which is used in (iv):
\begin{lstlisting}
m : ℕ
P : Fin m → ℕ
exp : Fin m → ℕ
hneq : ∏ i : Fin m, (P i) ^ (exp i) = n
hP : ∀ i, Nat.Prime (P i)
\end{lstlisting}

If the prime factors of $n$ are small (one or two digits), then a proof  \lstinline|hP| can be obtained automatically using the \lstinline|decide| tactic (we say more about primality proofs in Section \ref{seclpfw}).
The tuples $(a_i)_{i=0}^{n-1}$, $(b_i)_{i=0}^{n-1}$ are also fields in the structure: 
\begin{lstlisting}
a : Fin n → List (ZMod p)
b : Fin n → List (ZMod p)
\end{lstlisting}
As well as a proof of the fourth verification statement, written in terms of list arithmetic:
\begin{lstlisting}
hgcd : ∀ (i : Fin m),
(a ↑(n / P i) * L + b ↑(n / P i) * (h ↑(n / P i) - [0, 1]))
  .dropTrailingZeros = 1
\end{lstlisting}

The rest of the certifying data together with proofs of the verification statements is included in the structure in this way. Once a term of this type has been constructed, we get a proof that the polynomial \texttt{ofList L} is irreducible.

For an irreducible polynomial over $\FF_p$, a certificate of this form always exists. The function \texttt{CertificateIrrModpFFP}, implemented in SageMath, computes the certifying data --including the decomposition of the degree into prime factors-- and writes the corresponding term in Lean. For small degrees (usually $n \leq 12$) all of the proofs included as fields in this structure can be solved automatically either by using the \lstinline|rfl| or the \lstinline|decide| tactic.
For higher degrees, the computation may exceed Lean's preset maximum recursion depth or heartbeats limits. 
A variety of approaches could be used to deal with this timeout issue:
it might be possible to develop better decision procedures for use with \lstinline{decide},
or we might use (a custom extension to) the \lstinline{norm_num} tactic that performs this computation.

\subsection{Degree Analysis}\label{secdegree}
\looseness=-1
Here, we describe a certificate for the irreducibility of polynomials in $\ZZ[X]$, constructed by combining information from the factorization of the polynomial modulo distinct primes. 

Let $f$ in $\ZZ[X]$ of degree $n$ and $p$ a prime number not dividing $\operatorname{lc}(f)$, the leading coefficient of $f$. Let $\bar{f} = f \pmod p$ be the polynomial over $\FF_p$ obtained by reducing the coefficients of $f$ modulo $p$. Write $\bar{f} = \prod_{i = 1}^{m}g_i$, with $g_i$ irreducible in $\FF_p[X]$. 
 Consider the tuple $D_p = (\deg(g_i))_{i}$ and let $S(D_p)$ be the set of all possible subset sums of $D_p$. It is not hard to see that if $a$ divides $f$, then $\deg a \in S(D_p)$. Computing $S(D_p)$ for different primes $p$ will thus give us refined information on the degree of the factors of $f$. In particular:

\begin{proposition}\label{degreeanalysis}
Let $f$ in $\ZZ[X]$ be primitive and $p_1, \ldots, p_m$ a collection of distinct primes not dividing $\operatorname{lc}(f)$. Let 
\begin{align*}
\textstyle{ d = \min \left( \bigcap_{i = 1}^{m} S(D_{p_i}) \setminus \{0\}  \right)}.
\end{align*} If $a \in \ZZ[X]$ divides $f$ and is neither $1$ or $-1$, then $d \leq \deg a$. 
\end{proposition}
It follows that if $d = \deg f$, then $f$ is irreducible. This gives an irreducibility certification scheme, which
we formalized as the \lstinline|structure| \lstinline|IrreducibleCertificateIntPolynomial|. As parts of its fields, it includes data on the irreducible factors of $f \pmod p$ for different primes $p$. The proofs of irreducibility are handled using the certifying structure from Section \ref{finitefields}.

A certificate based on degree analysis, when it exists, is not unique. It is advisable to prioritize smaller primes that produce factors of small degrees for faster verification.  
Most polynomials (in an asymptotic sense) will have a certificate of this form, depending on their Galois group \cite{chebotarev}. However, there are cases where such a certificate does not exist.

\subsection{LPFW Certificate}\label{seclpfw}
\looseness=-1
This section describes an alternative certificate for irreducibility of polynomials in $\ZZ[X]$ which is more widely applicable. We discuss its formalization in Lean, and a script that can automatically generate a Lean proof of irreducibility based on these certificates. 

There is a way of reducing irreducibility testing in $\ZZ[X]$ to primality testing in $\ZZ$: 
if a `large enough' integer $m$ is found such that  $|f(m)|$ is prime, then $f$ can be shown to be irreducible. This test was 
proposed by Brillhart in \cite{Brillhart}. 
Abbott \cite{Abbott}, using a refined calculation that incorporates a known lower bound on the degrees of $f$'s factors, introduced the  \emph{large prime factor witness} (LPFW) certificate, which is similar to Brillhart's, but works for a larger number of cases and often produces smaller prime witnesses. It relies on the following proposition, which we have formalized. 
\begin{proposition}\label{lpfw}
Let $f$ be a non-constant polynomial in $\ZZ[X]$. Suppose that $d\in \ZZ_{\geq 1}$ is a known lower bound for the degree of the (non-unit) factors of $f$. Let $\rho$ in $\mathbb{R}$ be such that $|\alpha| \leq \rho$ for every root $\alpha \in \CC$ of $f$. If there exists $m \in \mathbb{Z}$ such that $|m| \geq \rho + 1$ and $|f(m)| = sP$ with an integer $s < (|m| - \rho) ^ d$ and $P \in \ZZ$ prime, then $f$ is irreducible. 
\end{proposition}

If a conjecture stated by Bouniakovsky \cite{bouniakowsky} is true, then there will always exist an LPFW certificate for any irreducible polynomial $f$ in $\ZZ[X]$. 
Ideally, the prime witness $P$ should be as small as possible so its primality can be quickly verified. The higher the lower bound $d$ for the degree of the factors of $f$, the more flexibility we have in finding such $P$. Note that $d$ can be obtained using Proposition \ref{degreeanalysis}. This leads us to combine the degree analysis techniques from the previous section with the aforementioned Proposition. 

To use Proposition \ref{lpfw}, we must find a root bound $\rho$. Classical bounds are given by Lagrange \cite{lagrange} and Cauchy \cite{cauchy}. 
For dense polynomials with integer coefficients, Cauchy's bound is typically sharper (unless $|a_n|$ is very large). By applying it to $f(rX)$, with $r>0$ a real number, one gets:

\begin{proposition}
Let $f = \sum_{i=0}^{n}a_i X^i $ in $\CC[X]$ of degree  $n>0$. Let $r > 0$ be a real number. If $\alpha \in \CC$ is a root of $f$, then 
$$ \textstyle{ |\alpha| \leq r \left ( 1 + \frac{1}{|a_n|} \max_{0 \leq i \leq n-1}  \frac{|a_i|}{r^{n-i} }  \right ) }$$
\end{proposition}

By setting $r=1$, we recover the original Cauchy bound. However, choosing the right $r$ can yield a sharper bound. We formalized this (scaled) Cauchy bound and proved: 
\begin{lstlisting}
lemma polynomial_roots_le_cauchy_bound_scale
  (f : Polynomial ℂ) (z : ℂ) (hd : f.natDegree ≠ 0)
  (hr : z ∈ f.roots) (r : ℝ) (hs : 0 < r) :
  Complex.abs z ≤ cauchyBoundScaled f r := ...
\end{lstlisting}
The term \lstinline|cauchyBoundScaled| 
takes as parameters a complex polynomial $f$ and real number $r$ and outputs a real number. However, since \mathlib polynomials are noncomputable and both complex and real numbers lack decidable equality, this definition is not optimal for computation. Thus, we defined a computable version, \lstinline|cauchyBoundScaledOfList|, for polynomials over $\ZZ$ (given as lists) using a rational scaling factor. 
This allows us to prove its value by computation:
\begin{lstlisting}
example : cauchyBoundScaledOfList [3,14,15,92,65] (1/2) =
  249/130 := by decide!
\end{lstlisting}
Recall from Section~\ref{sec:computation} that \lstinline|decide!| performs kernel reduction to prove decidable propositions.

We formalized the LPFW certification scheme as:
\begin{lstlisting}
structure CertificateIrreducibleIntOfPrimeDegreeAnalysis
  (f : Polynomial ℤ) (L : List ℤ)
\end{lstlisting}
It takes as parameters a polynomial $f$ over the integers and its coefficient list $L$. Similar to the certificate in the previous section, it includes fields that certify the factorization of $f$ modulo a given list of primes in order to obtain a lower bound $d$ on the degree of the factors of $f$, as in Proposition \ref{degreeanalysis}. The structure also includes fields such as the Cauchy bound $\rho$ --obtained using a scaling factor $r$--, the large prime witness $P$, the factor $s$, the evaluation argument $m$, as well as the proofs of the corresponding verification statements. 

One field in the structure is a proof of \lstinline|Nat.Prime P|. This is a decidable statement, and \mathlib provides the tactic \lstinline|norm_num| that can prove a number is prime if it is not too large ($< 7$ digits). However, the prime witness $P$ is typically big, so an alternative primality test might be necessary. Markus Himmel formalized Pratt certificates and a corresponding tactic \cite{pratt}, which can be used to prove such goals.

As before, irreducibility proofs for the factors of $f$ modulo primes can be handled as in Section \ref{finitefields}. All remaining proofs included as fields in the above structure can be solved automatically using \lstinline|decide|, \lstinline|decide!|, and/or \lstinline|norm_num|.  We also formalized a variant with the trivial degree lower bound $d=1$, resulting in a shorter certificate.

We wrote a SageMath script that outputs a Lean proof of the irreducibility of $f$ by selecting the appropriate certificate from Section \ref{secdegree} or Section \ref{seclpfw}. 
As noted in Section \ref{finitefields}, large polynomial degrees may cause issues with recursion depth or excesive heartbeats when checking the proofs in Lean. This approach of generating Lean code from SageMath worked well for our applications. However, exploring a tighter integration, such as calling SageMath directly from Lean, could be an interesting direction for future work. 

In the search for a smaller prime witness $P$, it can be helpful to transform $f$. Möbius transformations~\cite{Abbott}, and more generally, Tschirnhaus transformation~\cite[Algorithm 6.3.4]{Cohen}, can be used. Incorporating this into our proof generator is something we also intend to pursue in the future.

\section{Verifying Rings of Integers}\label{sec:rings-of-integers}
\looseness=-1
As stated in the introduction, our main goal is to formally verify that a ring $\OO$, with an explicitly given basis and contained inside a number field $K$, equals --in the set-theoretic sense-- the ring of integers of $K$. 
Computation in $\OO$ is key for both our proof and future applications. This raises two initial questions: how should we define $\OO$ in Lean, and how can we compute in it? 

\subsection{Construction of the Subalgebra}\label{subalgebrabuilder}
The goal of this section is to describe the construction of a ring $\mathcal{O} \subseteq K$ in Lean, given by an explicit basis. 

Consider a more general setting where $R$ and $K$ are commutative rings and $K$ is actually an $R$-algebra.
We recall that this means that there is a ring homomorphism $\phi: R \to K$ and as such $K$ carries the natural structure of an $R$-module by defining the scalar multiplication $r\cdot x$ (where $r \in R$ and $x \in K$) by $\phi(r)x$.
An $R$-subalgebra $\mathcal{O}$ of $K$ can be formalized in multiple ways. For example, we could consider a type \lstinline|(|$ \OO$ \lstinline|: Type*)|
with instances \lstinline|[CommRing| $\OO$\lstinline|]| \lstinline|[Algebra R| $\OO$ \lstinline|]| and equip it with an injective $R$-algebra homomorphism 
\lstinline|f :| $\OO$ \lstinline|→ₐ[R] K|. However, carrying along this map through constructions and proofs can be rather cumbersome. A better approach is to use the subalgebra \lstinline|structure| in \mathlib, which bundles together a subset of $K$, the carrier set, and the proofs that it is an $R$-algebra. In fact, the integral closure of $R$ in $K$ is defined in \mathlib with type \lstinline|Subalgebra R K|, allowing us to state our goal as \lstinline|⊢| $\OO$ \lstinline|= integralClosure R K|. 

Defining $\OO$ simply as the $R$-span of a given set gives a term of type \lstinline|Submodule R K|. To make this into a subalgebra we also need proofs that it is closed under multiplication and contains $1$. In order to automate this construction in Lean starting with an $R$-basis, we created the \texttt{SubalgebraBuilder}. 

To describe how it works, let us further specialize our setup and fix some notation. 
Suppose that $R$ is a domain, $Q$ its fraction field, and $K$ the extension of $Q$ obtained by adjoining a root $\theta$ of a monic polynomial $T$ in $R[X]$ of degree $n$ (e.g. $R=\ZZ$, $Q = \QQ$ and $K$ a number field). Suppose that $\{w_1, \ldots, w_n\}$ is an $R$-basis for a subalgebra $\OO$ of $K$.  Since $K \cong Q[X]/\langle T \rangle $, we can find a common denominator $d \in R\setminus\{0\}$ and polynomials $b_i \in R[X]$ of degree less than $n$ such that $w_i = \frac{1}{d}b_i(\theta)$.

As a module, $\OO$ is the $R$-span of $\{w_1, \ldots, w_n\}$. However, as previously remarked, in order to give it a type \lstinline|Subalgebra R K| instead of just \lstinline|Submodule R K|, we need to convince Lean that $\OO$ is closed under multiplication and contains $1$. 
To prove closure under multiplication it suffices to show that $w_i * w_j$ is in $\OO$ for all $i$ and $j$. A way to certify this is to give the coefficients $a_{ijk} \in R$ such that $w_i * w_j = \sum_{k} a_{ijk} \cdot w_k$. Verifying this equality in $K$ is equivalent to proving that for all $i$ and $j$ there exists a polynomial $s_{ij} \in R[X]$ such that 
\begin{align}\label{idenpolys}
\textstyle{b_i * b_j = d \sum_{k} a_{ijk} \cdot b_k- T * s_{ij}}.
\end{align}

Thus, the elements $a_{ijk}$ and polynomials $s_{ij}$ act as a certificate for the closure of $\mathcal{O}$ under multiplication. In fact, they provide a certificate for the precise representation of $w_i * w_j$ with respect to the basis $\{w_1, \ldots, w_n\}$. Since multiplication is commutative, verifying this certificate requires checking $(n^2+n)/2$ identities in $R[X]$. Proving that $1$ is in $\OO$ reduces to a single polynomial identity check. 

Besides constructing $\OO$ as a subalgebra in Lean, we also want a basis for it, which would be a term of type \mbox{\lstinline|Basis ι R|} $\OO$ (with \lstinline|ι| some indexing type). To prove that $\{w_1, \ldots, w_n\}$ is indeed a basis, we must show that it is linearly independent. Write $b_j = \sum_{i=0}^{n-1}c_{ij}X^i$ with $c_{ij} \in R$, then 
\begin{align*}
\textstyle {w_j = \frac{1}{d}\sum_{i=0}^{n-1}c_{ij}\theta^i}.
\end{align*}
The set $\{w_1, \ldots, w_n\}$ is linearly independent if and only if the determinant of the matrix $B = (c_{ij})$ is non-zero. To simplify this verification, we can \emph{choose} a basis where $B$ is upper triangular. In that case, it suffices to check that the diagonal entries are non-zero. 

Since any matrix over a Bézout domain has an echelon form \cite{linalg}, if $R$ is Bézout (e.g. $R = \ZZ$ or $R = \FF_p[X]$) there is a basis for $\OO$ with $B$ upper triangular. For $R = \ZZ$, 
a basis in Hermite normal form \cite[Theorem 4.7.3]{Cohen} may be used. 

We incorporate all this data and proofs into the structure: 
\begin{lstlisting}
structure SubalgebraBuilderLists
(n : ℕ) (R Q K : Type*) [CommRing R] [IsDomain R]
[DecidableEq R] [Field Q] [CommRing K] [Algebra Q K]
[Algebra R Q] [Algebra R K] [IsScalarTower R Q K]
[IsFractionRing R Q] (T : R[X]) (L : List R)
\end{lstlisting}
It takes as explicit parameters $R, Q, K$, the degree $n$, the monic polynomial $T$, and the list $L$ of coefficients of $T$.
We follow the practice in Mathlib to encode the structure of inclusion of
$R$ into its fraction field $Q$,
and $Q$ into the ring $K$ using the \lstinline{Algebra} and \lstinline{IsFractionRing} typeclasses.
The transitive inclusion of $R$ into $K$ is witnessed by \lstinline{[Algebra R K] [IsScalarTower R Q K]} instances;
the latter asserts the triangle of inclusions $R \to Q \to K$ commutes.
Browning and Lutz~\cite{berkeley-galois} explain \mathlib's approach to extensions (of fields) in depth,
and Wieser~\cite{mathlib-scalar-actions} covers scalar actions in \mathlib more generally.

The fields of the structure include the matrix $B^T$ --with $B$ upper triangular containing the basis coefficients $c_{ij}$ as described above--, the common denominator $d$, the coefficients $a_{ijk}$, the certifying polynomials $s_{ij}$ in list form, as well as the proofs that $B$ is upper triangular with non-zero elements in the diagonal, that the identities in  (\ref{idenpolys}) hold (using list-based arithmetic), and that $K$ is obtained by adjoining a root of $T$. 
All of these proofs, except the latter, can be solved automatically using the \lstinline|decide|, \lstinline|decide!|, and/or \lstinline|norm_num| tactics. The latter statement is conveyed with the predicate \lstinline|IsAdjoinRoot K (map (algebraMap R Q) T)|. Its proof will depend on the specific nature of $K$. If $K$ is \emph{defined} as \lstinline|AdjoinRoot (map (algebraMap R Q) T)|, then this proof is obtained for free. 

With $A$ of this type, the term
\begin{lstlisting}
subalgebraOfBuilderLists T L A : Subalgebra R K
\end{lstlisting}
is the corresponding $R$-subalgebra $\OO$ of $K$ which, as a set, is the $R$-span of $\{w_1, \ldots, w_n\}$, specified by $B$. We obtain a basis $(w_i)_{i = 1}^{n}$ for this subalgebra --indexed with \lstinline|Fin n|-- in \lstinline|basisOfBuilderLists T L A|, 
with $w_i = \frac{1}{d} b_i(\theta)$.

An element in $\OO$ can be represented by a vector \lstinline|Fin n → R| using its coordinates with respect to this basis. As previously noted, the polynomial identities in (\ref{idenpolys}) prove that the coordinates of $w_i * w_j$ are given by $(a_{ijk})_{k = 1}^{n}$. This information can be collected in a \textit{times table}: an $n \times n$ matrix where the $ij$-entry is the vector $(a_{ijk})_{k = 1}^n \in R^n$.  This notion is more generally formalized in Lean as a structure introduced in \cite{diophantine-toolkit-cpp} called \lstinline|TimesTable|, which bundles the data of a basis and the associated times table. 
\begin{lstlisting}
structure TimesTable ( (*$\iota$*) R S : Type*) [Semiring R]
    [AddCommMonoid S] [Mul S] [Module R S] : Type*
  where
basis : Basis ι R S
table : ι → ι → ι → R
basis_mul_basis : ∀ i j k, basis.repr (basis i * basis j) k = table i j k
\end{lstlisting}
Here, \lstinline|basis| is an $R$-basis for $S$, and \lstinline|basis.repr| sends an element in $S$ to its vector representation. The last field states that the product of the $i$-th and $j$-th basis elements is represented by the $ij$-th entry of \lstinline|table|. 
The \texttt{SubalgebraBuilder} contains the information to construct the times table for $\OO$. Indeed, \lstinline|timesTableOfSubalgebraBuilderLists T L A| has type
\lstinline|TimesTable (Fin n) R| $\OO$. 

\subsection{Arithmetic in $\OO$}\label{arithmetic}
We discuss how we can use a times table to compute in $\OO$. 

More generally, let $S$ be an $R$-algebra which is finite and free as an $R$-module and has basis $\mathcal{B} = \{w_1, \ldots, w_n\}$. An element $x$ in $S$ can be uniquely written as
$x = \sum_{i}c_i w_i$, and thus represented as a vector $(c_1, \ldots, c_n) \in R^n$. We have that $S \cong R ^ n$ as $R$-modules.  Furthermore, we can \emph{define} a multiplication on $R^n$ that corresponds to the multiplication in $S$:

Let $x = \sum_{i} c_i w_i$ and $y = \sum_{i} d_i w_i$, with $c_i, d_i \in R$, be elements in $S$. We have $xy = \sum_{i}\sum_{j} (c_id_j) \cdot(w_i * w_j)$. 
Let $T$ be the times table for $S$ with respect to $\mathcal{B}$, so that $T_{ij} = (a_{ijk})_{k} \in R ^ n$, with $w_i * w_j = \sum_{k} a_{ijk}w_k$. By defining
\begin{align}\label{mullists}
	\textstyle (c_i)_{i} * (d_i)_{i} := \sum_{i}\sum_{j}(c_id_j) \cdot T_{ij}
\end{align}
 in $R^n$, we can translate arithmetic in $S$ to arithmetic in $R^n$ and vice versa. If we have a times table for $S$, this approach allows us to compute in $S$ with Lean using lists as follows:

 Consider a $n \times n$ matrix $T$ with entries in $R^n$. In Lean, this is a term \lstinline|T : Fin n → Fin n → Fin n → R|.  Using our previously defined list arithmetic, we formalize (\ref{mullists}) as: 
 \begin{lstlisting}
def table_mul_list' (c d : List R) : List R :=
	List.sum (List.ofFn (fun i => List.sum (List.ofFn
	(fun j => List.mulPointwise ((List.getD c i 0) * (List.getD d j 0))  (List.ofFn (T i j))))))
 \end{lstlisting}
\vspace{-0.5cm}
This function is meant to be used with $c$ and $d$ of length $n$. However, we give them type \lstinline|List R| instead of \lstinline|Fin n → R| because lists offer faster computation. The term \lstinline|List.getD c i 0| retrieves the $i$-th entry of $c$, returning $0$ if $i$ is out of bounds. A variant where $T$ has type \lstinline|Fin n → Fin n → List R| is also defined for faster computation.

With this approach, we also recursively defined efficient exponentiation using the square-and-multiply method. Note that addition in $S$ corresponds to pointwise addition on lists.

We formally proved that, when $T$ is a times table for $S$, these functions on lists correspond to the arithmetic operations in $S$. 
Thus,  identities in $S$ --involving additions and multiplications-- can be proven by first translating them into their equivalent operations on lists. 
This method is used to prove identities in the subalgebra $\OO$ of a number field for which we have a times table \lstinline|TimesTable (Fin n) ℤ| $\OO$. This verification on lists of integers can be done automatically by a tactic such as \lstinline|decide!|. A similar strategy is used for the $\mathbb{F}_p$-algebra $\OO / p \OO$ from Section \ref{secnon}. 
\subsection{Local Maximality}
Here, we present a local notion of maximality which serves as a tool to prove an equality between subalgebras. 

Let $K$ be a number field of degree $n$ and $\mathcal{O}$ a subalgebra which is free and finite as a $\ZZ$-module. Then $\OO$ is contained in $\OO_K$, and we can compare them by considering the index $[\mathcal{O}_K : \mathcal{O}]$, typically defined as the cardinality of the quotient $\OO_K / \OO$. When $\mathcal{O}$ has rank $n$, then $[\mathcal{O}_K : \mathcal{O}]$ is a finite integer. 

If $p$ is a prime that does not divide $[\OO_K : \OO]$, we say that $\mathcal{O}$ is \emph{$p$-maximal}. Thus, if one can prove that $\OO$ is $p$-maximal for every prime $p$, it will follow that $\OO = \OO_K$. 

This idea, which will guide our approach for constructing a formal proof of $\OO = \OO_K$, is also central to the currently used algorithms for computing rings of integers. 
However, the approach of showing a \emph{global} equality by verifying a \emph{local} property is not limited to subrings in number fields.

We formalized a local notion of maximality that applies to free and finite modules over 
Principal Ideal Domains (abbreviated as \emph{PID}s, which means that every ideal can be generated by one element)
using a generalized notion of index, similar to \cite{lenstra}. This allows us to state and prove several useful results in wider generality.   
For $M$ a free module over a PID of finite rank, the index of a submodule $N$, which we also denote by $[M : N]$, is an element in $R$. If $[M : N]$ is a unit, then $M = N$. When $R = \ZZ$ and $M$ and $N$ have equal rank, this index coincides (up to sign) with the cardinality of $M/N$.

In Lean, to say that $R$ is a PID and $M$ is an $R$-module we use the hypothesis \lstinline|[CommRing R] [IsPrincipalIdealRing R]|
\newline \lstinline|[IsDomain R] [AddCommGroup M] [Module R M]|. We formalized the above notion of index as: 
\begin{lstlisting}
def Submodule.indexPID (N : Submodule R M) [Module.Free R M] [Module.Finite R M] : R
\end{lstlisting}
where we include \lstinline|[Module.Free R M] [Module.Finite R M]| as instance parameters, indicating that $M$ is free and finite. The definition is tagged \lstinline|noncomputable| as it depends on choice of bases, which are noncomputable. 

For an element $\pi$ in $R$, we say that $N$ is $\pi$-maximal if $\pi$ does not divide $[M : N]$. In formal terms, this reads 

\begin{lstlisting}
def piMaximal [Module.Free R M] [Module.Finite R M] 
  ((* $\pi$*) : R) (N : Submodule R M) : Prop :=
 	¬ ((* $\pi$*) ∣ Submodule.indexPID N)
\end{lstlisting}
From the properties of the index, it follows that if $N$ is $\pi$-maximal for every prime $\pi$, then $M = N$; where we recall that $\pi$ is prime means that it is a nonzero non-unit element such that $\pi | ab$ implies $\pi | a$ or $\pi | b$ for any $a,b \in R$.

As outlined at the start of this section, we will use this definition when $K$ is an $R$-algebra, $\OO$ and $\OO'$ have type \lstinline|Subalgebra R K|, and we have the hypothesis \lstinline|(hm : | $\OO \leq \OO' $\lstinline|)|.  
To state that $\OO$ is $\pi$-maximal (with respect to $\OO'$), we need to view $\OO$ as a submodule of $\OO'$. This is done via the inclusion map 
$\upmapsto \hspace{-0.1cm} \OO$ \lstinline|→ₐ[R] | $\upmapsto \hspace{-0.1cm} \OO'$. 
The term 
\begin{lstlisting}
Subalgebra.toSubmodule (Subalgebra.inclusion hm).range
\end{lstlisting}
is the range of this map, which we abbreviate as $\mathcal{O}*$, and has type
\lstinline|Submodule R| $\upmapsto \hspace{-0.1cm} \OO'$, where $\upmapsto \hspace{-0.1cm} \OO'$ is the coercion of $ \OO'$ into a subtype of $K$.

\subsection{The Pohst--Zassenhaus Theorem}
In this section, we introduce a crucial theorem that can be used to establish local maximality and forms the basis for the certificate discussed in Section \ref{secnon}. 

Let $R$ be a PID, $Q$ its fraction field, and $K$ an $R$-algebra. Consider two $R$-subalgebras of $K$, denoted by $\OO$ and  $\OO'$, which are free and finite as $R$-modules, have the same rank, and satisfy $\OO \subseteq \OO'$. The primary case of interest to keep in mind is: $R = \ZZ$, $K$ is a number field, and $\OO' = \OO_K$. 

To construct a proof of $\OO = \OO'$ we follow the strategy described in the previous section by establishing $\pi$-maximality for every prime $\pi$ of $R$. The main result we use is based on a theorem due to Pohst and Zassenhaus \cite[Lemma 5.53]{pohst}, which underpins the widely used \emph{Round-2 algorithm} for computing rings of integers \cite[Algorithm 6.1.8]{Cohen}. 

For $I$ and ideal of $\OO$, we define the \emph{multiplier ring} of $I$ by $r(I) = \{ x \in K \mid xI \subseteq I\}$. It is an $R$-subalgebra of $K$ which contains $\OO$. We formalized this definition as
\begin{lstlisting}
def multiplierRing (I : Ideal (*$\OO$*)) : Subalgebra R K where
	carrier :=  {(x : K)  | ∀ (i : (*$\OO$*)), i ∈ I → 
								( ∃ (j : (*$\OO$*)), j ∈ I ∧ i * x = j )} ...
\end{lstlisting}
The radical of $I$ is the ideal $\{x \in \OO \mid \exists m \geq 1 \text{,  }x^m \in I \}$. For $r \in R$, we denote by $I_{r}$ the radical of the principal ideal $r\OO$. The main theorem we use to prove $\pi$-maximality is: 
\begin{theorem}\label{pohstthm}
Let $\pi$ in $R$ be a prime and suppose that $\OO$ and $\OO'$ have equal rank. If $\OO = r(I_\pi)$, then $\OO$ is $\pi$-maximal. 
\end{theorem}
This is a version of the Pohst--Zassenhaus theorem which is sufficient for our purposes. However, it is more commonly presented with $K$ being a finite dimensional and separable $Q$-algebra, and $\OO'$ the integral closure of $R$ in $K$. In that case, it is an if-and-only-if statement \cite{Cohen}, \cite{pohst}. 

We formalized Theorem \ref{pohstthm} as follows: 
\begin{lstlisting}
variable (hm : (*$\OO$*) ≤ (*$\OO$*)') {(*$\pi$ *) : R} (hp : Prime (*$\pi$ *))
local notation "(*$\OO$*)*" => Subalgebra.toSubmodule  
		 (Subalgebra.inclusion hm).range
theorem order_piMaximal_of_order_eq_multiplierRing [Module.Free R (*$\OO$*)'] [Module.Finite R (*$\OO$*)']
 		 (heqr : Module.rank R (*$\OO$*) = Module.rank R (*$\OO$*)') 
 		 (heq : (*$\OO$*) = 
multiplierRing (Ideal.span {algebraMap R (*$\OO$*) π}).radical) :
	piMaximal π (*$\OO$*)*:= 
\end{lstlisting}

As before, $\OO*$ represents $\OO$ as a submodule of $\OO'$. The term \lstinline|Ideal.span {algebraMap R O π}| is the ideal $\pi \OO$, with \lstinline|algebraMap R| $\OO$ \lstinline|π| used to regard $\pi$ as an element of $\OO$. 

The proof strategy for this theorem, which we adapted from  \cite{Cohen}, consists of defining an intermediate subalgebra $\OO_\pi = \{x \in \OO' \mid \exists j \geq 0 \text{ such that } \pi ^ j  x \in \OO\}$, which sits between $\OO$ and $\OO'$ and is $\pi$-maximal by construction, and showing that $\OO = \OO_\pi$ whenever $\OO = r(I_\pi)$. 

Using Theorem \ref{pohstthm}, the proof of $\pi$-maximality of $\OO$ reduces to the verification of the equality $\OO = r(I_\pi)$. 
While an explicit description of $I_\pi$ may not be simple, in cases where $\OO \cong R[X]/\langle T \rangle$ --with $T$ a monic polynomial-- there is an explicit description of $I_\pi$ that leads to a simple criterion for $\pi$-maximality, called the Dedekind criterion. The case where $\OO$ is not necessarily of this form will be addressed in Section \ref{secnon} for $R = \ZZ$.

\subsection{Dedekind Criterion}\label{secdedekind}
We present a practical criterion involving only polynomial arithmetic that provides a simple way of establishing $\pi$-maximality in certain cases. 

Let $R$ be a PID, $Q$ its fraction field, and $\OO= R[X]/\langle T \rangle = R[\theta]$, with $\theta$ a root of $T$, a monic polynomial. For a prime $\pi \in R$, the quotient $R/\pi R$ is a field, and $(R/\pi R)[X]$ a PID.
To describe the ideal $I_\pi$ of $\OO$, we consider the reduction of $T$ modulo $\pi$. Denote by $T \pmod \pi$, or $\bar{T}$, the polynomial in $(R/\pi R)[X]$ obtained by mapping the coefficients of $T$ through the reduction map $R \to R/\pi R$. 
If $g \in R[X]$ is a lift of the radical of $\bar{T}$ --meaning that $\bar{g}$ is the product of the distinct irreducible factors of $\bar{T}$ -- then $I_\pi$ can be expressed as $I_\pi = \pi \OO + g(\theta) \OO$. This leads to a criterion that guarantees $\OO = r(I_\pi)$.

\begin{definition}\label{dedekind}
Let $g$ in $R[X]$ be a lift of the radical of $\bar{T}$. 
Set $h$ to be a lift of $\bar{T}/\bar{g}$ and define $f = (gh - T) / \pi$. 
We say that $T$ \emph{satisfies the Dedekind criterion} at $\pi$ if
$\gcd (\bar{f}, \bar{g}, \bar{h}) = 1$. 
\end{definition}

If $K$ is a torsion-free $R$-algebra containing $\OO$, then $T$ satisfying the Dedekind criterion at $\pi$ implies $\OO = r(I_\pi)$. Combined with Theorem \ref{pohstthm}, this leads to the proposition: 

\begin{proposition}\label{dedekindmaximal}
Let $K$ be torsion-free $R$-algebra containing $\OO \cong R[X]/\langle T \rangle $. Let $\OO'$ be an $R$-subalgebra of $K$, containing $\OO$ and with the same rank. If $T$ satisfies the Dedekind criterion at $\pi$, then $\OO$ is $\pi$-maximal (i.e. $\pi$ does not divide $[\OO' : \OO]$).  
\end{proposition}

To formalize the reduction modulo $\pi$ of polynomials as in Definition \ref{dedekind}, perhaps the obvious choice would be to use the quotient map from \lstinline|R| to \lstinline|R / (Ideal.span {|$\pi$\lstinline|})|. However, for verifying the Dedekind criterion we need to compute in $(R/\pi R)[X]$, and thus in $R/\pi R$. Hence, flexibility in the representation of $R/\pi R$ is important. For instance, we prefer to use polynomials over \lstinline|ZMod p| rather than \lstinline|ℤ / (Ideal.span {p})| as we have decidable equality and efficient modular arithmetic in the former type but not in the latter. 

For this reason, we model the quotient $R/\pi R$ as some type \lstinline|(F : Type*)| with \lstinline|[Field F]| instance, along with a ring homomorphism \lstinline|q : R →+* F | which is surjective and its kernel equals \lstinline|Ideal.span {|$\pi$\lstinline|}|. This ensures that $ F \cong R/\pi R$. 
This approach of capturing the essential relationship between structures without fixing a specific construction is common in \mathlib as it allows for greater flexibility. In this context, a proof transfer framework such as \cite{trocq} could also be useful for automating the translation of results involving different representations of $ R/\pi R$, though we did not explore this further. 

We formalized the Dedekind criterion as follows: 
\begin{lstlisting}
def satisfiesDedekindCriterion [Field F] (q : R →+* F) 
		((*$\pi$*) : R) (T : Polynomial R) : Prop :=
	∃ (f g h : Polynomial R) (a b c : Polynomial F),
	IsRadicalPart (g mod π ) (T mod π)
	∧ f * (C π) = g * h - T
	∧ (a * (f mod π) + b * (g mod π) + c * (h mod π) = 1)
	\end{lstlisting}
\vspace{-0.5cm}
The notation \lstinline|f mod π| stands for \lstinline|Polynomial.map q f|, which applies $q$ to the coefficients of $f$. This formulation of Definition \ref{dedekind} lets us avoid polynomial divisions and is more convenient, after providing the witness polynomials, for a proof by computation involving only simple operations. 
The proposition \lstinline|IsRadicalPart (g mod π ) (T mod π)| states that $\bar{g}$ and $\bar{T}$ have the same prime divisors and $\bar{g}$ is squarefree. We can prove it by verifying that $\bar{g} \mid \bar{T}$, that $\bar{T} \mid \bar{g}^n$ for some $n$, and that $\bar{g}$ is squarefree. 
The latter can be done by checking that $\gcd(\bar{g},\bar{g}') = 1$, with $\bar{g}'$ the formal derivative of $\bar{g}$ . 

Observe that the discussed assumptions on \lstinline|q : R →+* F| are not needed to write down the previous definition. However, they do appear in the following theorem, which is the formal version of Proposition \ref{dedekindmaximal}, adapted from \cite{Cohen}: 
\begin{lstlisting}
variable (q : R →+* F) (hqsurj : Function.Surjective q)
(hqker : RingHom.ker q = Ideal.span {(*$\pi$*)}) (hmc : (*$\OO$*) ≤ (*$\OO$*)') 
local notation "(*$\OO$*)*" => Subalgebra.toSubmodule  (Subalgebra.inclusion hmc).range 
theorem piMaximal_of_satisfiesDedekindCriteria 
		[Module.Free R (*$\OO$*)'] [Module.Finite R (*$\OO$*)']
		(j: IsAdjoinRoot (*$\OO$*) T) (hp : Prime π ) (hm: T.Monic)
		(heqr : Module.rank R (*$\OO$*)= Module.rank R (*$\OO$*)') 
		(h : satisfiesDedekindCriterion q π T) :
	piMaximal π (*$\OO$*)* := ...
\end{lstlisting}
Note that no assumption of irreducibility of $T$ is required. 

We also defined a specialized version of the Dedekind criterion for the case $R = \ZZ$ and \lstinline|F = ZMod p|. 
We created a \lstinline|structure|, called \lstinline|CertificateDedekindCriterionLists|, with parameters \lstinline| (T : ℤ[X]) (p : ℕ)|. Its fields include data certifying that $T$ satisfies the Dedekind criterion at $p$, and proofs of the associated identities --involving additions and multiplications and using our list-based approach-- which can be solved automatically by \lstinline|rfl| or \lstinline|decide|. 

The Dedekind criterion can still be useful even if $\OO$ is not given by adjoining a root of a monic polynomial. If $\alpha \in \mathcal{O}$ has minimal polynomial $T \in R[X]$, then the $R$-subalgebra $R[\alpha] \subseteq \OO$ is isomorphic to $R[X]/\langle T \rangle$.  If $T$ satisfies the Dedekind criterion at $\pi$ and $\OO'$ has rank equal to $ \deg(T)$, then $R[\alpha]$, and hence $\OO$, is $\pi$-maximal.  
Yet, for a number field $K$, the hope of using the Dedekind criterion to prove $p$-maximality of its ring of integers at every prime $p$ soon fades. While this criterion always applies if $\OO_K$ is monogenic,  in some number fields the index $[\OO_K : \ZZ[\alpha]]$ is divisible by a fixed prime for all $\alpha \in \OO_K$. Dedekind gave an example of this \cite{Dedekind}. 

\subsection{The Non-monogenic Case}\label{secnon}

In this section, we introduce a certificate for local maximality which can be used even if the Dedekind criterion does not apply. We specialize to $R = \ZZ$, our main interest. Our formalization efforts were primarily focused on this case, with $\FF_p$ modeled as  \lstinline|ZMod p|. However, we formalized many theorems about semilinear maps (introduced in Lean in~\cite{semilinear}) and their kernels --which we use extensively-- in a more general setting, laying the groundwork for future extensions. 

The strategy used in the \emph{Round-2 algorithm} is to reduce the computation of $I_p$ and $r(I_p)$ to a computation in linear algebra over the field $\FF_p$. 
We will use an analogous strategy to certify $p$-maximality by verifying that the kernel of certain linear map is trivial. 
Let $\OO$ be a commutative ring which is free as a $\ZZ$-module and of finite rank $r$, and let $p$ be a prime number. The quotient ring $\OO / p\OO$ is a finite dimensional $\FF_p$-algebra.
Similarly, the quotient $I_p / p I_p$ is a finite dimensional $\FF_p$-module (albeit not an $\FF_{p}$- algebra). 
Consider the map from $\OO/p\OO$ to the $\FF_p$-linear endomorphisms of $I_p/pI_p$.
\begin{align}
	\varphi : \OO / p\OO &\to \operatorname{End}(I_p / pI_p) \label{map_end} \\
	\bar{\alpha} &\mapsto (\bar{\beta} \to \overline{\alpha\beta}) \notag
\end{align} 
This is the map in \cite[Lemma 6.1.7]{Cohen}, it is well defined and $\FF_p$-linear. The following proposition follows directly from the definitions of the multiplier ring and the radical $I_p$ and will be the basis for our certificate of $p$-maximality. 

\begin{proposition}\label{propker}
Let $K$ be $\ZZ$-torsion-free ring containing $\OO$. If the kernel of the map $\varphi$ in (\ref{map_end}) is trivial, then $r(I_p) = \OO$. 
\end{proposition}

When $K$ is a $\QQ$-algebra, this is an if-and-only-if statement. If
$\OO'$ is a subring of $K$ containing $\OO$ and of equal rank, then the previous proposition together with Theorem (\ref{pohstthm}) ensures that $\OO$ is $p$-maximal if the kernel of $\varphi$ is trivial. Therefore, finding a way to certify that $\varphi$ has trivial kernel becomes our main goal.

To define (\ref{map_end}) in Lean, we need to endow both $I_p /pI_p$ and $\OO/p\OO$ with an $\FF_p$-module structure. However, these quotient constructions are not exactly the same. 
If we want Lean to automatically infer that  $\OO/p\OO$ has a \lstinline|CommRing| instance, the quotient $\OO/p\OO$ should be constructed with $p\OO$ of type \lstinline|Ideal| $\OO$ (which is definitionally equal to \lstinline|Submodule| $\OO$ $\OO$). However, since $I_p$ is not a ring (it lacks a one), $pI_p$ cannot be treated as an ideal of $I_p$.

To unify these two constructions, we model $M/nM$ for any abelian additive group $M$ and a non-zero natural $n$, by using a ring $R$ as a parameter (such that $M$ is an $R$-module) and considering the quotient as a quotient of $R$-modules. We represent it as
 \mbox{\lstinline|M / (n : R) • (⊤ : Submodule R M )|}, where \lstinline|⊤| is $M$ viewed as a submodule of itself. 
We then define: 
\begin{lstlisting}
	instance module_modp_is_zmodp_module 
		(R M : Type*) (n : ℕ)[Module R M] [NeZero n] :
	Module (ZMod n) (M / (n : R) • ⊤) :=
\end{lstlisting}
Thus, $I_p/pI_p$ (taking $M= I_p$ and $R = \ZZ$) and $O/pO$ (taking $M = \OO$ and $R = \OO$) are both special cases of this construction, allowing Lean to infer an $\FF_p$-module instance for each. Furthermore, Lean will automatically synthesize a \lstinline|CommRing| instance for $\OO / p\OO$, which we then use to give it a $\FF_p$-algebra structure. 
Another advantage of this unified approach is that we can define associated objects --such as a $\ZZ/n\ZZ$-basis for $M/nM$ given a $\ZZ$-basis for $M$-- and apply it to both cases. 
We also provide \lstinline{DistribMulAction} and \lstinline{SMulCommClass} instances that act as a shortcut
for typeclass inference, and thereby prevent a timeout in downstream code.

To define map (\ref{map_end}) in Lean, we used multiple steps. First, for a parameter \lstinline|(α :| $\OO$\lstinline|)|, we defined the map $I_p \to I_p/pI_p$ sending $\beta \mapsto \overline{\alpha\beta}$. Using \lstinline|Quotient.lift| we lifted this to a map $I_p/pI_p \to I_p/pI_p$. This was bundled with its proofs of linearity into a linear function, giving an element in $\operatorname{End}(I_p / pI_p)$. After defining the corresponding map $\OO \to \operatorname{End}(I_p / pI_p)$, lifting again, and proving linearity, we obtain:

\begin{lstlisting}
def map_to_end_lin ((*$\OO$*) : Type*) [CommRing (*$\OO$*)]
(p : N) [Fact (Nat.Prime p)] :
(*$\OO$*)  / ↑p • ⊤ →(*\textsubscript{l}*)[ZMod p]  (*$\upmapsto \hspace{-0.1cm}$*) (Ideal.radical (↑p • ⊤)) / ↑p • ⊤ 
→(*\textsubscript{l}*)[ZMod p] (*$\upmapsto \hspace{-0.1cm}$*) (Ideal.radical (↑p • ⊤)) / ↑p • ⊤ := ...
\end{lstlisting}

The arrows have the following meanings. The expression \lstinline|↑p| appearing in $\OO$\lstinline| / ↑p • ⊤| and \lstinline|Ideal.radical (↑p • ⊤)| denotes the coercion of \lstinline|p : ℕ| into $\OO$. The arrow $\upmapsto $ coerces \lstinline|Ideal.radical (↑p • ⊤)| into a type. Lastly, the rightmost appearance of \lstinline|↑p| in \lstinline|(Ideal.radical (↑p • ⊤)) / ↑p • ⊤| is the coercion of \lstinline|p : ℕ| into \lstinline|ℤ|. 

Formally, Proposition \ref{propker} reads: 
\begin{lstlisting}
theorem mult_ring_eq_ring_of_trivial_ker_map_to_end_lin 
 	{K : Type*} [CommRing K] [NoZeroSMulDivisors ℤ K] 
 	((*$\OO$*) : Subalgebra ℤ K) (p : ℕ) [hpI : Fact (Nat.Prime p)] 
	(hk : LinearMap.ker (map_to_end_lin (*$\upmapsto \hspace{-0.1cm} \OO$*) p) = ⊥) :
(*$\OO$*) = multiplierRing (Ideal.span {↑p}).radical := 
\end{lstlisting}

Let us return to the matter of certifying that the kernel of $\varphi$ is trivial. Given a $\ZZ$-basis for an abelian additive group $M$, we obtain a $\ZZ/n\ZZ$-basis for $M/nM$ by mapping it through the $\sigma$-semilinear map $M \to M/nM$, with $\sigma : \ZZ \to \ZZ/n\ZZ$. With these bases, the image of an element in $M$ represented by $(x_1, \ldots, x_r) \in \ZZ^r$ will have coordinates $(\bar{x}_1, \ldots, \bar{x}_r) \in (\ZZ/n\ZZ)^r$ in $M/nM$. 
 In this way, we obtain an $\FF_p$-basis for $\OO / p\OO$ from a $\ZZ$-basis for $\OO$. 
Finding an $\FF_p$-basis for $I_p/pI_p$ is more complex. We will make use of the Frobenius endomorphism $\Fr_p$, which, in a ring of characteristic $p$, is the $\FF_p$-linear map $x \mapsto x^p$. For $t \in \NN$, the \emph{iterated} Frobenius endomorphism $\Fr_p^t$ sends $x \mapsto x^{p^t}$. 
The image of $I_p$ under the quotient map $\OO \to \OO/p\OO$ is $I_p / p\OO$, an $\FF_p$-subspace of $\OO / p\OO$. The following proposition lets us realize $I_p / p\OO$ as the kernel of a linear map:
\begin{proposition}
Let $t$ be an integer such that $r \leq p ^ t$. The kernel of $\Fr_p ^t : \OO / p\OO \to \OO / p\OO$ is equal to $I_p / p\OO$. 
\end{proposition}
While we could theoretically find a matrix representation for this map, an algorithm for computing the kernel of a matrix has not yet, as far as we know, been implemented in Lean. However, we can \emph{certify} that a given set of elements in $\OO/p\OO$ is a basis for the kernel of $\Fr_p^t$ as follows:

Denote by $\bar{x}$ the image in $\OO/p\OO$ of $x$ in $\OO$. Provide multisets $\mathcal{V} = \{v_1, \ldots, v_m\}$ and $\mathcal{W} = \{w_1, \ldots, w_n\}$ with elements in $\OO$ --where $r = m + n$-- such that $\overline{\mathcal{V}} = \{\bar{v}_1, \ldots, \bar{v}_m\}$ and $\mathcal{U}=\{\Fr_p^t(\bar{w}_1), \ldots, \Fr_p^t(\bar{w}_n)\}$ are $\FF_p$-linearly independent and $\Fr_p^t(\bar{v}_i) = 0$ for all $i$. By a dimension argument, this guarantees that $\{\bar{v}_1, \ldots, \bar{v}_m\}$ is an $\FF_p$-basis for the kernel of $\Fr_p^t$ -- and thus for $I_p/p\OO$.

We can represent $\overline{\mathcal{V}}$ as an $m \times r$ matrix over $\FF_p$, the rows being the coordinates of the elements in $\overline{\mathcal{V}}$ with respect to the basis for $\OO/p\OO$. 
Since we are working over a field, we can \emph{choose} the elements in $\mathcal{V}$ so that the corresponding 
matrix is in reduced row echelon form. The linear independence of $\overline{\mathcal{V}}$ can then be read off directly, without any determinant calculation (which is rather slow using the current Lean implementation). Similarly, the right choice of $\mathcal{W}$ will make the linear independence of $\mathcal{U}$ immediate to determine. 

With the $v_i$ and $w_i$ as above, it can be shown that 
\begin{align}\label{abasis}
\{\bar{v}_1, \ldots, \bar{v}_m, \overline{pw}_1, \ldots, \overline{pw}_n\} \subseteq I_p/pI_p
\end{align}
is an $\FF_p$-basis for $I_p/pI_p$, with $\bar{y}$ the image of $y \in I_p$ in $I_p/pI_p$.

Many different quotients and reductions come into play in this last statement, this made our initial formalization attempt quite challenging. To address this, we proved it in a more general setting as \lstinline|basisSubmoduleModOfBasisMod|. 
Instead of using the quotient construction, we worked with general types endowed with the appropiate instances, incorporating surjective maps with conditions on their kernels. This approach also allowed the construction to type-check faster. 
The result is a term of type \lstinline|Basis (Fin m ⊕ Fin n) S J | (in this case, $J$ models $I_p/pI_p$ and $S$ corresponds to $\FF_p$). 

Using \lstinline|Fin m ⊕ Fin n| as the index type, which is the disjoint union of \lstinline|Fin m | and \lstinline|Fin n|, mimics our informal description and makes it more convenient to work with.

With a basis for $I_p/pI_p$, we can now represent an element in this space as an $r$-tuple,  and an element in $\operatorname{End}(I_p/p_p)$ as an $r \times r$ matrix over $\FF_p$. To \emph{certify} that the kernel of the map $\varphi$ in (\ref{map_end}) is trivial, we can provide a multiset $\mathcal{X} = \{x_1, \ldots, x_r\}$ of elements in $\OO$ such that $\varphi(\overline{\mathcal{X}}) = \{\varphi(\bar{x_1}), \ldots, \varphi(\bar{x_r})\}$ is linearly independent. This last verification can be simplified by observing that, generically, we expect to find an element $\beta_W \in I_p/pI_p$  such that 
$ \{\varphi(\bar{x}_1)(\beta_W), \ldots, \varphi(\bar{x}_r)(\beta_W)\}$ is linearly independent. This $\beta_W$ acts as a witness of the linear independence of $\varphi(\overline{\mathcal{X}})$. Furthermore, we can \emph{choose} $\mathcal{X}$ in a way that makes the linear independence of the multiset easy to verify. This leads us to the following certificate for $p$-maximality.

\paragraph{Certificate} Consider $\OO$  and $\OO'$ subalgebras of a $\ZZ$-torsion-free commutative ring such that $\OO \subseteq \OO'$, and both of rank $r$. Let $\mathcal{B} = \{b_1, \ldots, b_r\}$ be a $\ZZ$-basis for $\OO$. Let $m,n$ and $t$ be non-negative integers with $r = m + n$ and $r \leq p^t$.
A certificate for $p$-maximality consists of the following data: 
\noindent
\begin{minipage}[t]{0.25\textwidth}
	\begin{itemize}
		\item $m \times r$ matrix $\mathcal{V}$ over $\ZZ$
		\item $n \times r$ matrix $\mathcal{W}$ over $\ZZ$
		\item $n \times r$ matrix $\mathcal{U}$ over $\FF_p$
		\item $m$-tuple $\nu$ over $\ZZ_{>0}$
		\item $n$-tuple $\omega$ over $\ZZ_{>0}$
	\end{itemize}
\end{minipage}%
\hspace{-0.5cm}
\begin{minipage}[t]{0.55\textwidth}
	\begin{itemize}
		\item $r\times r$ matrix $\mathcal{X}$ over $\ZZ$
		\item $m$-tuple $\beta$ over $\ZZ$
		\item $n$-tuple $\gamma$ over $\ZZ$ 
		\item $r\times m$ matrix $\mathbf{a}$ over $\ZZ$ 
		\item $r\times n$ matrix $\mathbf{c}$ over $\ZZ$ 
	\end{itemize}
\end{minipage}
\begin{itemize}
	\item $r$-tuple $\eta$ over the disjoint union $A\sqcup B$ with\\ $A = \{1, \ldots, m\}$ and $B = \{1, \ldots, n\}$.
\end{itemize}
For an integer $s$, let $\bar{s}$ be its reduction modulo $p$ in $\FF_p$. For $b$ in $\OO$, let $\bar{b}$ be its reduction in $\OO/p\OO$. 
For verification, check that:

\begin{enumerate}[label=(\roman*)]
	\item For all $i$, $\overline{\mathcal{V}}_{i\nu_i} \neq 0$ and $\overline{\mathcal{V}}_{j\nu_i} = 0$ for all $j \neq i$.
	\item For all $i$, $\mathcal{U}_{i\omega_i} \neq 0$ and $\mathcal{U}_{j\omega_i} = 0$ for all $j \neq i$.
	\item For all $i$, 
	\begin{itemize}[left = -3pt]
		\item if $\eta_i$ is in $A$: $\overline{a_{i\eta_i}} \neq 0$ and $\forall j \neq i$, $\overline{a_{j\eta_i}} = 0$; 
		\item if $\eta_i$ is in $B$: $\overline{c_{i\eta_i}} \neq 0$ and $\forall j \neq i$, $\overline{c_{j\eta_i}} = 0$. 
	\end{itemize}
	\item For all $i$, $(\sum_{j} \overline{\mathcal{V}_{ij}} \cdot \overline{b_j})^{p^t} = 0$.
	\item For all $i$, $(\sum_{j} \overline{\mathcal{W}_{ij}} \cdot \overline{b_j})^{p^t} = \sum_{j} \mathcal{U}_{ij} \cdot \overline{b_j}$. 
	\item For all $i$ :
	\begin{multline*}
		\textstyle  (\sum_{j} \mathcal{X}_{ij} \cdot b_j) (\sum_{k} \left ( \beta_{k} \cdot \sum_{j} \mathcal{V}_{kj} \cdot b_j \right ) + \sum_{k} \left ( p\gamma_{k} \cdot \sum_{j} \mathcal{W}_{kj} \cdot b_j \right ))\\ =\textstyle \sum_{k} \left ( a_{ik} \cdot \sum_{j} \mathcal{V}_{kj} \cdot b_j \right ) + \sum_{k} \left ( pc_{ik} \cdot \sum_{j} \mathcal{W}_{kj} \cdot b_j \right ). 
	\end{multline*}
\end{enumerate} 

The rows of $\mathcal{X}$, $\mathcal{V}$ and $\mathcal{W}$ represent elements in $\OO$ as remarked before. The matrix $\mathcal{U}$ represents elements in $\OO/p\OO$. Statement (i) and (ii) prove that $\overline {\mathcal{V}}$ and $\mathcal{U}$ are linearly independent. This, together with (iv) and (v) guarantee that the collection as in (\ref{abasis}) is a basis for $I_p/pI_p$.
The tuples $\gamma$ and $\beta$ give the coordinates of a witness element $\beta_W$ in $I_p/pI_p$ with respect to this basis. Statement (vi) together with (iii) imply that 
$ \{\varphi(\bar{x}_1)(\beta_W), \ldots, \varphi(\bar{x}_r)(\beta_W)\}$ is linearly independent, establishing the $p$-maximality of $\OO$. 

We formalized this certification scheme in Lean as:
\begin{lstlisting}
structure MaximalOrderCertificateWLists 
		{K : Type*} [CommRing K] [NoZeroSMulDivisors ℤ K]
		(p : ℕ) [hpI : Fact $ Nat.Prime p] ((*$\OO$*) : Subalgebra ℤ K) ((*$\OO$*)' : Subalgebra ℤ K) (hm : (*$\OO$*) ≤ (*$\OO$*)') where ... 
\end{lstlisting}

It includes as part of its fields a term of type \lstinline|TimesTable (Fin (m + n)) ℤ| $\OO$, a term of type \lstinline|Basis (Fin (m + n)) ℤ| $ \OO'$ (which we do not know explicitly, of course, but serves as a proof that, as a $\ZZ$-module, $\OO'$ is free, finite and has rank $r$), 
and the certifying data described earlier, together with proofs of the verification statements. 

The statements (iv), (v), and (vi), which involve $r$ identities in $\OO/p\OO$ and $r$ in $\OO$, are stated in terms of lists over $\FF_p$ and $\ZZ$, respectively, as described in Section \ref{arithmetic}. The proofs can then be solved using \lstinline|decide| and/or \lstinline|decide!|. 
From a term of this type, we get a proof of \lstinline|piMaximal ↑p| $\OO$ \hspace{-0.15cm} \lstinline|*|. 

In case a witness $\beta_W$ does not exist or cannot be found, we formalized a longer certification scheme for which a certificate always exists when $K$ is a number field, $\OO' = \mathcal{O}_K$, and $\OO$ is $p$-maximal. This requires verifying $r^2$ identities in $\OO$. For details, we refer to Appendix~\ref{appendix:certificate}. Additionally, in cases where $m = 0$ (corresponding to $p$ being \emph{unramified} in $K$) a simplified certificate is available. 

We wrote  \texttt{CertificatePMaximalityF} in SageMath, to compute these certificates and write the corresponding Lean term. 

\subsection{Global Maximality}
Here, we describe how we can automatically generate a Lean proof of $\OO = \OO_K$ using the certificates previously introduced. 

Let $K$ be the number field obtained by adjoining a root $\theta$ of the monic irreducible polynomial $T$ in $\ZZ[X]$. We aim to verify that a subalgebra $\OO$, with an explicitly given basis, is equal to $\OO_K$. 
Since $T$ is separable, we can find $a ,b \in \ZZ[X]$ such that $aT + bT'= n$ for some nonzero integer $n$.  
If $p$ does not divide $n$, then $T \pmod p$ is squarefree and it easily follows that $T$ satisfies the Dedekind criterion at $p$. 
 Given $\theta \in \OO$, this gives a quick way to prove $p$-maximality of $\OO$ for \emph{all} but finitely many primes $p$ (those dividing $n$). In fact, $T$ may still satisfy the criterion for some of these primes. For the rest, the certificates in Section \ref{secnon} can be used, leading to a proof of $\OO = \OO_K$. 

We wrote a SageMath function, called \texttt{LeanProof}, that takes as input an irreducible and monic polynomial $T \in \ZZ[X]$, a $\ZZ$-basis $\{w_i\}_i$ for the ring of integers of $K = \QQ[X]/\langle T \rangle $, and outputs a Lean proof of :
\begin{lstlisting}
(*$\OO$*) =  NumberField.RingOfIntegers K
\end{lstlisting}
where $K$ is \lstinline|AdjoinRoot (map (algebraMap ℤ ℚ) T) |, and $\OO$ is the subalgebra of $K$ with integral basis $\{w_i\}_i$, constructed as in Section \ref{subalgebrabuilder}. The proof of irreducibility of $T$ is generated as in Section \ref{sec:irreducibility}. Lean can then infer a \lstinline|NumberField K| instance. The term \lstinline|NumberField.RingOfIntegers K| is simply the coercion of \lstinline|IntegralClosure ℤ K| into a type. As mentioned, while generating Lean code from SageMath was sufficient for our purposes, tighter integration between the CAS and Lean remains an interesting direction for future work.

The previous approach avoids any computation of the discriminant (defined in the next section) in Lean. However, a proof of the discriminant of $\OO$, and consequently of $K$, can help narrow down the primes one needs to check. These concept, and its computation, are the subject of the next section.

\section{Resultants and Discriminants}\label{sec:resultants-discriminants}
The goal of this section is to provide a method for computing the discriminant of an integral basis of $\OO$.
Consider an $R$-algebra $A$ with a finite $R$-basis. Given an element $x \in A$, the map $y \mapsto xy$ is an $R$-linear map, and thus can be represented as a matrix with entries in $R$. The  trace of this matrix, which is independent of the choice of basis, is known as the \emph{trace} of $x$ and denoted by $\operatorname{Tr}_{A/R}(x)$. For a collection of elements $a_1, \ldots, a_n$ of $A$, its \emph{discriminant} is defined by $\operatorname{disc}(a_1, \ldots, a_n) = \det(\operatorname{Tr}_{A/R}(a_i a_j))_{ij}$. When we started our project, \mathlib already contained this definition. 

If $K = \QQ(\alpha)$ is a number field of degree $n$ and $\OO$ is a subring which is free of rank $n$ as a $\ZZ$-module, we define $\operatorname{disc}(\OO) \in \ZZ$ as the discriminant of a $\ZZ$-basis for $\OO$. In particular, $\operatorname{disc}(\OO_K)$ is known as the \emph{discriminant} of the number field $K$ and it is an important arithmetic invariant encoding information about $K$. 
If $\alpha \in \OO$, then we have 
\begin{align}\label{discrK}
\operatorname{disc}(1, \alpha, \ldots, \alpha^{n-1}) = [\mathcal{O} : \ZZ[\alpha]]^2 \cdot \operatorname{disc}(\OO) .
\end{align}
The index $[\mathcal{O} : \ZZ[\alpha]]$ can be determined from an explicit basis for $\OO$ in terms of $\alpha$. Knowing  the discriminant of the power basis $1, \alpha, \ldots, \alpha ^{n-1}$ then allows us to compute $\operatorname{disc(\OO)}$.

More generally, let $K = F(\alpha)$ be a finite separable field extension of $F$ (e.g. $K$ a number field and $F = \QQ$). Letting the conjugate roots of $\alpha$ in the algebraic closure be numbered $\alpha_1, \dots, \alpha_n$,
then the discriminant of $1, \alpha, \ldots, \alpha^{n-1}$ is equal to $\prod_i \prod_{j > i} (\alpha_j - \alpha_i)^2$. 
This result was already available in \mathlib. However, the value of the discriminant is not directly computable. Thus, we define a computable notion of discriminant for an explicitly given polynomial.

In the remainder of this section, let $R$ be an integral domain,
$f, g$ be polynomials of degree $n$ and $m$ respectively with coefficients given by
$f(X) = a_n X^n + \cdots + a_1 X + a_0 \in R[X]$ and $g(X) = b_m X^m + \cdots + b_1 X + b_0 \in R[X]$.
Our path towards formalized computation of the discriminant starts with the \emph{resultant} $\Res(f, g)$
defined as the determinant of the \emph{Sylvester matrix} of dimension $(m + n) \times (m + n)$, determined by the coefficients of $f$ and $g$ (see \cite[IV, \S 8]{Lang} for a transposed version with $f$ and $g$ interchanged and coefficients reversed);
we work out an example in Appendix~\ref{appendix:sylvester-computation}.
Let $f$ be a polynomial with formal derivative $f'$,
then $\Res(f, f')$ is divisible by the leading coefficient $a_n$ of $f$;
one sees this by inspection of the last row of the Sylvester matrix,
which contains two nonzero entries: $a_n$ and $n a_n$.
We define the discriminant of a polynomial $f$ with leading coefficient $a_n$
such that $a_n \Disc(f) = (-1)^{n(n - 1)/2} \Res(f, f')$,
by modifying the Sylvester matrix to divide the last row by $a_n$,
and multiplying the determinant of this modified matrix by the desired sign.

Our definition of the resultant allows for direct evaluation,
but we need the relation to the discriminant of a power basis. 
The main theorem we proved expresses the resultant of two polynomials in terms of their roots:
\begin{theorem} \label{thm:resultantProd}
	Let $f, g \in R[X]$ be polynomials and assume they split completely: $f(X) = a_n \prod_i (X - \alpha_i)$ and $g(X) = b_m \prod_i (X - \beta_i)$,
	then:
	\begin{align*}
	\Res(f, g) = a_n^{m} b_m^{n} \prod_i \prod_j (\alpha_i - \beta_j).
	\end{align*}
\end{theorem}
Note that the assumption of Theorem~\ref{thm:resultantProd} always applies after mapping $f$ and $g$ to the algebraic closure of the fraction field of $R$,
and this map will preserve the resultant,
so we will assume that $f$ and $g$ split completely. It then follows from the product rule and elementary algebraic manipulations that, for $f$ monic with roots $\alpha_i$, 
$\prod_i \prod_{j > i} (\alpha_j - \alpha_i)^2 = \Disc(f)$.
Since the former term is shown in \mathlib to equal the discriminant of the power basis $1, \ldots, \alpha ^{n-1}$, and the latter term is computable, this gives us a way to compute, given a basis for $\OO_K$ and using (\ref{discrK}), the discriminant of a number field.

It remains to prove Theorem~\ref{thm:resultantProd}.
We first showed that for two homogeneous polynomials $p, q \in R[x, y, \dots, z]$ of the same degree:
if $p$ divides $q$, then they are equal up to a multiplicative constant: $q = c p$ for $c \in R$.
This required formalizing a substantial amount of preliminaries in the theory of (homogeneous) multivariate polynomials.

Next, we may take $f$ and $g$ monic and view $\Res(f, g)$ and $\prod_i \prod_j (\alpha_i - \beta_j)$ as polynomials in $R[\alpha_1, \dots, \alpha_n, \beta_1, \dots, \beta_m]$.
We showed that these are indeed homogeneous polynomials, of degree $m n$, where an essential ingredient was characterizing the homogeneous polynomials $p \in R[x, y, \dots, z]$ as those polynomials where $p(c x, c y, \dots, c z) = c^k p(x, y, \dots, z)$, assuming $R$ is infinite.
The coefficient of $(\alpha_1 \alpha_2 \cdots \alpha_n)^m$ in $\Res(f, g)$ and in $\prod_i \prod_j (\alpha_i - \beta_j)$ equals $1$,
so if the two expressions are equal up to multiplication by $c$, we must have $c = 1$ and the expressions are actually equal.

Finally, our approach to showing $\prod_i \prod_j (\alpha_i - \beta_j)$ divides $\Res(f, g)$
was by showing $\alpha_i - \beta_j$ divides $\Res(f, g)$ for each $i, j$.
We proved that the latter statement is equivalent, under the condition that $R$ is infinite,
to the statement that $\Res(f, g) = 0$ if $f$ and $g$ share a root.
To show the latter statement, we considered the $R$-linear map
$(p, q) \mapsto (p f + q g)$ on the subspace $M = \{(p, q) \in R[X] \times R[X] \mid \deg p < m \text{ and} \deg q < n\}$,
and computed that the determinant of this map is exactly equal to the resultant.
If $f$ and $g$ have a common root, then $(p, q) \mapsto (p f + q g)$ has a nontrivial kernel,
so the determinant is zero.

Although this approach does not differ significantly from one found in textbooks such as Lang's Algebra~\cite{Lang},
a particular source of difficulty we had to deal with was the textbooks' implicitly shifting point of view between different polynomial rings.
For example, Lang argues that $\Res(f, g)$ is homogeneous of degree $m$ as a polynomial in $R[a_0, \dots, a_n]$ and of degree $n$ as a polynomial in $R[b_0, \dots, b_m]$,
but that $\alpha_i - \beta_j$ divides $\Res(f, g)$ in $R[\alpha_1, \dots, \alpha_n, \beta_1, \dots, \beta_m]$.
Rather than having to repeat the boilerplate involved with a change of rings repeatedly throughout our proofs,
we chose to adapt the proofs from the literature to remain within one ring.
Our formalization works exclusively in the polynomials \newline $R[\alpha_1, \dots, \alpha_n, \beta_1, \dots, \beta_m]$.
For example, polynomials that are expressed in terms of $a_0, \dots, a_n, b_0, \dots, b_m$ can be translated to $R[\alpha_1, \dots, \alpha_n, \beta_1, \dots, \beta_m]$ using the monadic \lstinline{bind} operator~\cite{Witt-vectors},
since each coefficient $a_i$ can be given as a polynomial in the roots $\alpha_j$.

To evaluate discriminants, we use the \lstinline|Polynomial.ofList| definition of Section~\ref{sec:asLists}.
Expressions such as \lstinline|resultant (ofList [5, 4, 3, 2, 1]) (ofList [4, 6, 6, 4])| are computable,
although we run into a subsequent error: the implementation of determinants in \mathlib is not optimized for computation
and therefore causes a stack overflow.
However, we were still able to compute the discriminant of degree $3$ polynomials such as $f  = X^3 - 30X - 80$
and, after formally verifying an integral basis for the ring of integers of $K = \QQ[X]/\langle f \rangle$, proved
\begin{lstlisting}
theorem K_discr : NumberField.discr K = -16200 := ... 
\end{lstlisting}

\section{Formally Verifiying LMFDB Entries}\label{sec:lmfdb}

A popular publicly available online database containing number theoretic data is the \emph{L-functions and modular forms database} (LMFDB)~\cite{lmfdb}.
It gathers data from various sources, relying on several programming languages and CASes.
Much of the data is currently beyond the reach of formalization, but our work makes it possible to formally verify some nontrivial entries for \emph{number fields} of arbitrary degree.
In particular, we focus on checking the \emph{integral basis} for a choice of number fields up to degree $8$.
To a lesser extent, we also compute the corresponding \emph{discriminant}, but only up to degree $3$, as our computational methods are currently still under development.
At this point the aim is not to formally verify integral bases and discriminants for a large fraction of the LMFDB, but to show the feasibility, and current computational limits, of the approach. 

Concretely, there are $142$ number fields in the LMFDB which are of degree $5$, unramified outside $2,3,5$, and have (to focus on complications) non-monogenic ring of integers. (This list, up to isomorphism, is complete, but this is not our concern here.)
For all these $142$ number fields we successfully used our tools to formally verify the given \emph{integral basis}. The most time-consuming step when checking these proofs appears to be the exponentiation step in the certificates of Section \ref{secnon}. We recall that this is done with repeated squaring and multiplication.
Using a virtual machine running an AMD EPYC 7B13 processor (2.45 GHz) and 16GB of RAM, each ring of integers took, on average, around 33 seconds to check. Approximately 14\% of that time is spent on the irreducibility proof for the defining polynomial. Replacing \lstinline|decide| and \lstinline|decide!| with \lstinline|native_decide| whenever possible reduces the overall checking time by about 20\%.  During development, we encountered a couple of examples that required more than the default number of maximum heartbeats (200,000) for type-checking, and the proofs had to be manually adjusted. This was resolved after a \mathlib update. 

As our discriminant computations cannot yet handle large degrees, we also tested degree $3$ cases. There are $7$ number fields in the LMFDB which are of degree $3$, unramified outside $2,3,5$, and have non-monogenic ring of integers.
For all these number fields we successfully formally verify the given \emph{integral basis} as well as the \emph{discriminant}. We note that in all of these cases, the longer form of the certificate for $p$-maximality (given in detail in Appendix~\ref{appendix:certificate}) had to be used. In contrast, for all of degree 5 examples except one, the shorter certificate with a witness was sufficient. In this case, each ring of integers took, on average, around 28 seconds to check. Using \lstinline|native_decide| decreases this time by around 10\%.

We also successfully verified a degree 6 example and a simpler degree 8 example. Both took around 48 seconds to check. For these, \lstinline|native_decide| reduces the checking time by 61\% and 32\%, respectively. 


\section{Discussion}\label{sec:discussion}

\subsection{Future Work}

The discriminant computations should be improved.
At this point, these are directly computed from the corresponding resultant.
For repeated use of discriminant evaluation of a fixed degree, general formulae for the discriminant of a degree $n$ polynomial should be formalized, up to some \lq reasonable\rq\ $n$.
The computations for the ring of integers are in a more advanced stage, though further optimization is needed to deal with higher degrees in order to be able to cover all number fields from the LMFDB. We note that much of the mathematical foundation is formalized in enough generality to allow for possible extension to more general contexts. 
Although not essential for database verification -- since all necessary Lean files can be generated in advance-- integrating SageMath and Lean, for instance by calling our SageMath scripts directly from Lean, could be convenient for on-demand computations. 
Finally, certifying other fundamental invariants, such as class groups and unit groups, would be very interesting.
While these would constitute completely new projects, they could build on the formalized rings of integers.

\subsection{Related Work}

As mentioned above, some rings of integers for quadratic number fields have been formalized before~\cite{diophantine-toolkit-cpp}; this concerns both the determination as well as concrete computation.
Specific instances, such as the Gaussian integers $\ZZ[i]$, have been considered in various systems; see e.g.~\cite{gaussian_integers-isabelle, gaussian_integers-mizar}. 
Mathematical certificates outside of algebraic number theory have received much attention.
Some examples relevant for (general) number theory and computer algebra include~\cite{Thiemann-et-al-20, mahboubi-sibut-pinote-2021}. Previous work bridging Lean with Computer Algebra Systems includes the tactics \lstinline|polyrith|, written by Dhruv Bhatia, and \lstinline|sageify|~\cite{diophantine-toolkit-cpp}, both of which integrate with SageMath. Additionally, ~\cite{Lewis_2017, Lewis_bi_directional} develop a connection between Lean and Mathematica. 
The use of a proof assistant as a CAS itself has also been explored in prior work. 
In \cite{maxime}, efficient algorithms in linear algebra and polynomial arithmetic are implemented and verified in Coq. Similarly, \cite{aransay, aransay2} implements various linear algebra algorithms in Isabelle/HOL. 

\subsection{Conclusions}

We have successfully built tools to formally verify rings of integers in number fields of, in principle, arbitrary degree, as well as set up the basis for computing discriminants. This represents a step towards bridging the gap between traditional Computer Algebra Systems and formal verification. 
Developing some of the actual certification schemes, especially for the ring of integers, required several new ideas.
While the formalization of the mathematical theory proved challenging at times, a major part of the effort involved translating the abstract mathematical concepts and results into a representation suitable for computation. 

In total, for our developments, we wrote about 13000 lines of Lean code.
Our SageMath scripts automatically generated about
32000 lines of Lean code (excluding the alternative code containing \lstinline{native_decide}) to perform the certifications we considered.

With these kind of projects, it is very hard to reliably measure the De Bruijn factor.

\appendix
\section{Certificate for $p$-maximality}\label{appendix:certificate}

The following certificate is guaranteed to always exist if $K$ is a number field, $\OO' = \OO_K$, and $\OO$ is $p$-maximal. 
 
\subsection*{Certificate} Consider $\OO$  and $\OO'$ subalgebras of a $\ZZ$-torsion-free commutative ring such that $\OO \subseteq \OO'$, and both of rank $r$. Let $\mathcal{B} = \{b_1, \ldots, b_r\}$ be a $\ZZ$-basis for $\OO$. Let $m,n$ and $t$ be non-negative integers with $r = m + n$ and $r \leq p^t$.
A certificate for $p$-maximality consists of the following data: 

\noindent
\begin{minipage}[t]{0.25\textwidth}
	\begin{itemize}
		\item $m \times r$ matrix $\mathcal{V}$ over $\ZZ$
		\item $n \times r$ matrix $\mathcal{W}$ over $\ZZ$
		\item $n \times r$ matrix $\mathcal{U}$ over $\FF_p$
		\item $m$-tuple $\nu$ over $\ZZ_{>0}$
		\item $n$-tuple $\omega$ over $\ZZ_{>0}$
	\end{itemize}
\end{minipage}%
\hspace{-0.5cm}
\begin{minipage}[t]{0.55\textwidth}
	\begin{itemize}
		\item $r\times r$ matrix $\mathcal{X}$ over $\ZZ$
		\item $r\times m \times m$ array $\mathbf{a}$ over $\ZZ$. 
		\item $r\times m \times n$ array $\mathbf{c}$ over $\ZZ$. 
		\item $r\times n \times m$ array $\mathbf{d}$ over $\ZZ$. 
		\item $r\times n \times n$ array $\mathbf{e}$ over $\ZZ$. 
	\end{itemize}
\end{minipage}
\begin{itemize}
	\item $r$ pairs $(\eta_i, \eta'_i)$ with $\eta'_i$ and $ \eta_i$ in the disjoint union $A\sqcup B$ with $A = \{1, \ldots, m\}$ and $B = \{1, \ldots, n\}$
\end{itemize}
For an integer $s$, let $\bar{s}$ be its reduction modulo $p$ in $\FF_p$. For $b$ in $\OO$, let $\bar{b}$ be its reduction in $\OO/p\OO$. 
For verification, check that:

\begin{enumerate}[label=(\roman*)]
	\item For all $i$, $\overline{\mathcal{V}}_{i\nu_i} \neq 0$ and $\overline{\mathcal{V}}_{j\nu_i} = 0$ for all $j \neq i$
	\item For all $i$, $\mathcal{U}_{i\omega_i} \neq 0$ and $\mathcal{U}_{j\omega_i} = 0$ for all $j \neq i$
	\item For all $i$, 
	
	\begin{itemize}[left = -3pt]
		\item If $\eta_i$ and $\eta'_i$ in $A$, $\overline{a_{i\eta_i\eta'_i}} \neq 0$ and  $\forall j \neq i$, $\overline{a_{j\eta_i\eta'_i}} = 0$. 
		\item If $\eta_i$ in $A$ and $\eta'_i$ in $B$, $\overline{c_{i\eta_i\eta'_i}} \neq 0$ and  $\forall j \neq i$, $\overline{c_{j\eta_i\eta'_i}} = 0$. 
		\item If $\eta_i$ in $B$ and $\eta'_i$ in $A$, $\overline{d_{i\eta_i\eta'_i}} \neq 0$ and  $\forall j \neq i$, $\overline{d_{j\eta_i\eta'_i}} = 0$. 
		\item If $\eta_i$ and $\eta'_i$ in $B$, $\overline{e_{i\eta_i\eta'_i}} \neq 0$ and  $\forall j \neq i$, $\overline{e_{j\eta_i\eta'_i}} = 0$. 
	\end{itemize}
	\item For all $i$, $(\sum_{j} \overline{\mathcal{V}_{ij}} \cdot \overline{b_j})^{p^t} = 0$
	\item For all $i$, $(\sum_{j} \overline{\mathcal{W}_{ij}} \cdot \overline{b_j})^{p^t} = \sum_{j} \mathcal{U}_{ij} \cdot \overline{b_j}$. 
	\item For all $i$ and $j$ :
	\begin{multline*}
		\textstyle  (\sum_{l} \mathcal{X}_{il} \cdot b_l) (\sum_{l} \mathcal{V}_{jl}\cdot  b_l) = \\ 	\textstyle \sum_{k} \left ( a_{ijk} \cdot \sum_{l} \mathcal{V}_{kl} \cdot b_l \right ) + \sum_{k} \left ( pc_{ijk} \cdot \sum_{l} \mathcal{W}_{kl} \cdot b_l \right ). 
	\end{multline*}
	\item For all $i$ and $j$: 
	\begin{multline*}
		\textstyle  (\sum_{l} \mathcal{X}_{il} \cdot b_l) (\sum_{l} p\mathcal{W}_{jl}\cdot  b_l) = \\ 	\textstyle \sum_{k} \left ( d_{ijk} \cdot \sum_{l} \mathcal{V}_{kl} \cdot b_l \right ) + \sum_{k} \left ( pe_{ijk} \cdot \sum_{l} \mathcal{W}_{kl} \cdot b_l \right ).
	\end{multline*}
\end{enumerate} 

The rows of $\mathcal{X}$, $\mathcal{V}$ and $\mathcal{W}$ represent elements in $\OO$. The matrix $\mathcal{U}$ represents elements in $\OO/p\OO$. Statement (i) and (ii) prove that $\overline {\mathcal{V}}$ and $\mathcal{U}$ are linearly independent. This, together with (iv) and (v) guarantee that the collection as in (\ref{abasis}) is a basis for $I_p/pI_p$. Statements (vi) and (vii) imply that the matrix representing the endomorphism $\varphi(\bar{x_i})$ with respect to this basis is given by: 
\begin{align}\label{matrix}
	\left (
	\begin{smallmatrix}
		\overline{a_{i11}} & \dots & \overline{a_{im1}} & \overline{d_{i11}} & \dots & \overline{d_{in1}} \\
		\vdots & \ddots & \vdots & \vdots & \ddots & \vdots \\
		\overline{a_{i1m}} & \dots & \overline{a_{imm}} & \overline{d_{i1m}} & \dots & \overline{d_{inm}} \\
		\overline{c_{i11}} & \dots & \overline{c_{im1}} & \overline{e_{i11}} & \dots & \overline{e_{in1}} \\
		\vdots & \ddots & \vdots & \vdots & \ddots & \vdots \\
		\overline{c_{i1n}} & \dots & \overline{c_{imn}} & \overline{e_{i1n}} & \dots & \overline{e_{inn}}. 
	\end{smallmatrix} \right )
\end{align}
From (iii) the linear independence of $\{\varphi(\bar{x_1}), \ldots, \varphi(\bar{x_r}))\}$ follows, establishing the $p$-maximality of $\OO$. 

We formalized this certificate in Lean as the structure
\begin{lstlisting}
	structure MaximalOrderCertificateLists 
	{K : Type*} [CommRing K] [NoZeroSMulDivisors ℤ K]
	(p : ℕ) [hpI : Fact $ Nat.Prime p] ((*$\OO$*) : Subalgebra ℤ K) ((*$\OO$*)' : Subalgebra ℤ K) (hm : (*$\OO$*) ≤ (*$\OO$*)') where ... 
\end{lstlisting}

\section{Sylvester matrix computation examples}\label{appendix:sylvester-computation}

Consider the polynomial $f = 2 X^3 - X^2 - 2X + 1 \in \ZZ[X]$, which has derivative $f' = 6X^2 - 2X - 2$.
The Sylvester matrix for $f$ and $f'$ is given by:
\begin{align*}
	A &= \begin{pmatrix}
		-2 &  0 &  0 &  1 &  0 \\
		-2 &  -2 &  0 & -2 &  1 \\
		6 & -2 &  -2 & -1 & -2 \\
		0 & 6 & -2 &  2 & -1 \\
		0 &  0 & 6 &  0 &  2
	\end{pmatrix}.
\end{align*}

The resultant is defined as the determinant of the Sylvester matrix, giving us $\Res(f, f') = |A| = -72$.

To compute the discriminant of $f$ we modify the Sylvester matrix by dividing out the leading coefficient of $f$, $2$, in the last row:
\begin{align*}
	A &= \begin{pmatrix}
		-2 &  0 &  0 &  1 &  0 \\
		-2 &  -2 &  0 & -2 &  1 \\
		6 & -2 &  -2 & -1 & -2 \\
		0 & 6 & -2 &  2 & -1 \\
		0 &  0 & 3 &  0 &  1
	\end{pmatrix}.
\end{align*}

We obtain $\Disc(f) = (-1)^{\deg f (\deg f - 1)/2} |A'| = (-1)^3 \cdot -36 = 36$.

Passing to the field of fractions, we can divide $f$ by its leading coefficient to get the monic polynomial
$g = X^3 - \frac{1}{2} X^2 - X + \frac{1}{2} \in \QQ[X]$,
with discriminant computed as before $\Disc(g) = \frac{9}{4} = \left(\frac{1}{2}\right)^{2 (\deg f - 1)} \Disc(f)$.
The roots of $f$ and $g$ in $\QQ$ are given by $X \in \{-1, \frac{1}{2}, 1\}$,
and we find as expected that the product of root differences equals the discriminant of $g$:
\begin{align*}
	\left(\frac{1}{2} - -1\right)^2 \left(1 - -1\right)^2 \left(1 - \frac{1}{2}\right)^2 = \frac{9}{4}.
\end{align*}

\section*{Data-Availability Statement}
Full source code of our formalization and SageMath scripts are available online at \url{https://github.com/alainchmt/RingOfIntegersProject}
and are archived as~\cite{RingOfIntegersProject-archive}.

\section*{Acknowledgments}

Alain Chavarri Villarello was funded by NWO Vidi grant No. 613.009.143, Formalizing Diophantine algorithms.
We thank Assia Mahboubi, Filippo A. E. Nuccio and the anonymous referees for their helpful comments.

\bibliographystyle{ACM-Reference-Format}
\bibliography{references.bib}


\begin{thebibliography}{45}


\ifx \showCODEN    \undefined \def \showCODEN     #1{\unskip}     \fi
\ifx \showDOI      \undefined \def \showDOI       #1{#1}\fi
\ifx \showISBNx    \undefined \def \showISBNx     #1{\unskip}     \fi
\ifx \showISBNxiii \undefined \def \showISBNxiii  #1{\unskip}     \fi
\ifx \showISSN     \undefined \def \showISSN      #1{\unskip}     \fi
\ifx \showLCCN     \undefined \def \showLCCN      #1{\unskip}     \fi
\ifx \shownote     \undefined \def \shownote      #1{#1}          \fi
\ifx \showarticletitle \undefined \def \showarticletitle #1{#1}   \fi
\ifx \showURL      \undefined \def \showURL       {\relax}        \fi
\providecommand\bibfield[2]{#2}
\providecommand\bibinfo[2]{#2}
\providecommand\natexlab[1]{#1}
\providecommand\showeprint[2][]{arXiv:#2}

\bibitem[Abbott(2020)]%
        {Abbott}
\bibfield{author}{\bibinfo{person}{John Abbott}.} \bibinfo{year}{[2020]
  \copyright 2020}\natexlab{}.
\newblock \showarticletitle{Certifying irreducibility in {$\Bbb Z[x]$}}.
\newblock In \bibinfo{booktitle}{\emph{Mathematical software---{ICMS} 2020}}.
  \bibinfo{series}{Lecture Notes in Comput. Sci.},
  Vol.~\bibinfo{volume}{12097}. \bibinfo{publisher}{Springer, Cham},
  \bibinfo{pages}{462--472}.
\newblock
\showISBNx{978-3-030-52200-1; 978-3-030-52199-8}
\urldef\tempurl%
\url{https://doi.org/10.1007/978-3-030-52200-1\_46}
\showDOI{\tempurl}


\bibitem[Aransay and Divas\'on(2016)]%
        {aransay2}
\bibfield{author}{\bibinfo{person}{Jes\'us Aransay} {and} \bibinfo{person}{Jose
  Divas\'on}.} \bibinfo{year}{2016}\natexlab{}.
\newblock \showarticletitle{Formalisation of the computation of the echelon
  form of a matrix in {I}sabelle/{HOL}}.
\newblock \bibinfo{journal}{\emph{Form. Asp. Comput.}} \bibinfo{volume}{28},
  \bibinfo{number}{6} (\bibinfo{year}{2016}), \bibinfo{pages}{1005--1026}.
\newblock
\showISSN{0934-5043,1433-299X}
\urldef\tempurl%
\url{https://doi.org/10.1007/s00165-016-0383-1}
\showDOI{\tempurl}


\bibitem[Aransay and Divas\'on(2017)]%
        {aransay}
\bibfield{author}{\bibinfo{person}{Jes\'us Aransay} {and} \bibinfo{person}{Jose
  Divas\'on}.} \bibinfo{year}{2017}\natexlab{}.
\newblock \showarticletitle{A formalisation in {HOL} of the fundamental theorem
  of linear algebra and its application to the solution of the least squares
  problem}.
\newblock \bibinfo{journal}{\emph{J. Automat. Reason.}} \bibinfo{volume}{58},
  \bibinfo{number}{4} (\bibinfo{year}{2017}), \bibinfo{pages}{509--535}.
\newblock
\showISSN{0168-7433,1573-0670}
\urldef\tempurl%
\url{https://doi.org/10.1007/s10817-016-9379-z}
\showDOI{\tempurl}


\bibitem[Baanen(2020)]%
        {baanenring}
\bibfield{author}{\bibinfo{person}{Anne Baanen}.} \bibinfo{year}{[2020]
  \copyright 2020}\natexlab{}.
\newblock \showarticletitle{A lean tactic for normalising ring expressions with
  exponents (short paper)}.
\newblock In \bibinfo{booktitle}{\emph{Automated reasoning. {P}art {II}}}.
  \bibinfo{series}{Lecture Notes in Comput. Sci.},
  Vol.~\bibinfo{volume}{12167}. \bibinfo{publisher}{Springer, Cham},
  \bibinfo{pages}{21--27}.
\newblock
\showISBNx{978-3-030-51054-1; 978-3-030-51053-4}
\urldef\tempurl%
\url{https://doi.org/10.1007/978-3-030-51054-1\_2}
\showDOI{\tempurl}


\bibitem[Baanen et~al\mbox{.}(2023)]%
        {diophantine-toolkit-cpp}
\bibfield{author}{\bibinfo{person}{Anne Baanen}, \bibinfo{person}{Alex~J.
  Best}, \bibinfo{person}{Nirvana Coppola}, {and} \bibinfo{person}{Sander~R.
  Dahmen}.} \bibinfo{year}{2023}\natexlab{}.
\newblock \showarticletitle{Formalized Class Group Computations and Integral
  Points on Mordell Elliptic Curves}. In \bibinfo{booktitle}{\emph{{CPP}
  2023}}, \bibfield{editor}{\bibinfo{person}{Robbert Krebbers},
  \bibinfo{person}{Dmitriy Traytel}, \bibinfo{person}{Brigitte Pientka}, {and}
  \bibinfo{person}{Steve Zdancewic}} (Eds.). \bibinfo{publisher}{{ACM}},
  \bibinfo{pages}{47--62}.
\newblock
\urldef\tempurl%
\url{https://doi.org/10.1145/3573105.3575682}
\showDOI{\tempurl}


\bibitem[Baanen et~al\mbox{.}(2022)]%
        {class-group-jar}
\bibfield{author}{\bibinfo{person}{Anne Baanen}, \bibinfo{person}{Sander~R.
  Dahmen}, \bibinfo{person}{Ashvni Narayanan}, {and} \bibinfo{person}{Filippo
  A. E. Nuccio Mortarino~Majno di Capriglio}.} \bibinfo{year}{2022}\natexlab{}.
\newblock \showarticletitle{A Formalization of {D}edekind Domains and Class
  Groups of Global Fields}.
\newblock \bibinfo{journal}{\emph{J. Autom. Reason.}} \bibinfo{volume}{66},
  \bibinfo{number}{4} (\bibinfo{year}{2022}), \bibinfo{pages}{611--637}.
\newblock
\urldef\tempurl%
\url{https://doi.org/10.1007/s10817-022-09644-0}
\showDOI{\tempurl}


\bibitem[Bosma et~al\mbox{.}(1997)]%
        {Magma}
\bibfield{author}{\bibinfo{person}{Wieb Bosma}, \bibinfo{person}{John Cannon},
  {and} \bibinfo{person}{Catherine Playoust}.} \bibinfo{year}{1997}\natexlab{}.
\newblock \showarticletitle{The {M}agma algebra system. {I}. {T}he user
  language}.
\newblock \bibinfo{journal}{\emph{J. Symbolic Comput.}} \bibinfo{volume}{24},
  \bibinfo{number}{3-4} (\bibinfo{year}{1997}), \bibinfo{pages}{235--265}.
\newblock
\showISSN{0747-7171}
\urldef\tempurl%
\url{https://doi.org/10.1006/jsco.1996.0125}
\showDOI{\tempurl}
\newblock
\shownote{Computational algebra and number theory (London, 1993)}.


\bibitem[Bouniakowsky(1854)]%
        {bouniakowsky}
\bibfield{author}{\bibinfo{person}{Viktor Bouniakowsky}.}
  \bibinfo{year}{1854}\natexlab{}.
\newblock \bibinfo{booktitle}{\emph{Sur les diviseurs num{\'e}riques
  invariables des fonctions rationnelles entieres}}.
\newblock \bibinfo{publisher}{De l'Imprimerie de l'Acad{\'e}mie imp{\'e}riale
  des sciences}.
\newblock


\bibitem[Boutin(1997)]%
        {reflection-tactics}
\bibfield{author}{\bibinfo{person}{Samuel Boutin}.}
  \bibinfo{year}{1997}\natexlab{}.
\newblock \showarticletitle{Using reflection to build efficient and certified
  decision procedures}. In \bibinfo{booktitle}{\emph{TACS 1997}}
  \emph{(\bibinfo{series}{LNCS}, Vol.~\bibinfo{volume}{1281})},
  \bibfield{editor}{\bibinfo{person}{Mart{\'i}n Abadi} {and}
  \bibinfo{person}{Takayasu Ito}} (Eds.). \bibinfo{publisher}{Springer, Berlin,
  Heidelberg}, \bibinfo{pages}{515--529}.
\newblock
\showISBNx{978-3-540-69530-1}
\urldef\tempurl%
\url{https://doi.org/10.1007/11541868_7}
\showDOI{\tempurl}


\bibitem[Brillhart(1980)]%
        {Brillhart}
\bibfield{author}{\bibinfo{person}{John Brillhart}.}
  \bibinfo{year}{1980}\natexlab{}.
\newblock \showarticletitle{Note on irreducibility testing}.
\newblock \bibinfo{journal}{\emph{Math. Comp.}} \bibinfo{volume}{35},
  \bibinfo{number}{152} (\bibinfo{year}{1980}), \bibinfo{pages}{1379--1381}.
\newblock
\showISSN{0025-5718,1088-6842}
\urldef\tempurl%
\url{https://doi.org/10.2307/2006403}
\showDOI{\tempurl}


\bibitem[Browning and Lutz(2022)]%
        {berkeley-galois}
\bibfield{author}{\bibinfo{person}{Thomas Browning} {and}
  \bibinfo{person}{Patrick Lutz}.} \bibinfo{year}{2022}\natexlab{}.
\newblock \showarticletitle{Formalizing Galois Theory}.
\newblock \bibinfo{journal}{\emph{Exp. Math.}} \bibinfo{volume}{31},
  \bibinfo{number}{2} (\bibinfo{year}{2022}), \bibinfo{pages}{413--424}.
\newblock
\urldef\tempurl%
\url{https://doi.org/10.1080/10586458.2021.1986176}
\showDOI{\tempurl}


\bibitem[Buchmann and Lenstra(1994)]%
        {lenstra}
\bibfield{author}{\bibinfo{person}{J.~A. Buchmann} {and} \bibinfo{person}{H.~W.
  Lenstra, Jr.}} \bibinfo{year}{1994}\natexlab{}.
\newblock \showarticletitle{Approximating rings of integers in number fields}.
\newblock \bibinfo{journal}{\emph{J. Th\'eor. Nombres Bordeaux}}
  \bibinfo{volume}{6}, \bibinfo{number}{2} (\bibinfo{year}{1994}),
  \bibinfo{pages}{221--260}.
\newblock
\showISSN{1246-7405,2118-8572}
\urldef\tempurl%
\url{http://jtnb.cedram.org/item?id=JTNB_1994__6_2_221_0}
\showURL{%
\tempurl}


\bibitem[Carneiro(2019)]%
        {lean-tt}
\bibfield{author}{\bibinfo{person}{Mario Carneiro}.}
  \bibinfo{year}{2019}\natexlab{}.
\newblock \bibinfo{title}{The Type Theory of {L}ean}.
\newblock
\newblock
\urldef\tempurl%
\url{https://github.com/digama0/lean-type-theory/releases}
\showURL{%
\tempurl}
\newblock
\shownote{MSc thesis}.


\bibitem[Cauchy(1829)]%
        {cauchy}
\bibfield{author}{\bibinfo{person}{Augustin Louis~Baron Cauchy}.}
  \bibinfo{year}{1829}\natexlab{}.
\newblock \bibinfo{booktitle}{\emph{Exercices de math{\'e}matiques}}.
  \bibinfo{series}{{\OE}uvres 2}, Vol.~\bibinfo{volume}{9}.
\newblock 122 pages.
\newblock


\bibitem[Chavarri~Villarello et~al\mbox{.}(2024)]%
        {RingOfIntegersProject-archive}
\bibfield{author}{\bibinfo{person}{Alain Chavarri~Villarello},
  \bibinfo{person}{Sander Dahmen}, {and} \bibinfo{person}{Anne Baanen}.}
  \bibinfo{year}{2024}\natexlab{}.
\newblock \bibinfo{booktitle}{\emph{Certifying rings of integers in number
  fields}}.
\newblock
\urldef\tempurl%
\url{https://doi.org/10.5281/zenodo.14283856}
\showDOI{\tempurl}


\bibitem[Cohen et~al\mbox{.}(2024)]%
        {trocq}
\bibfield{author}{\bibinfo{person}{Cyril Cohen}, \bibinfo{person}{Enzo Crance},
  {and} \bibinfo{person}{Assia Mahboubi}.} \bibinfo{year}{2024}\natexlab{}.
\newblock \showarticletitle{Trocq: proof transfer for free, with or without
  univalence}. In \bibinfo{booktitle}{\emph{European Symposium on
  Programming}}. Springer, \bibinfo{pages}{239--268}.
\newblock


\bibitem[Cohen(1993)]%
        {Cohen}
\bibfield{author}{\bibinfo{person}{Henri Cohen}.}
  \bibinfo{year}{1993}\natexlab{}.
\newblock \bibinfo{booktitle}{\emph{A course in computational algebraic number
  theory}}. \bibinfo{series}{Graduate Texts in Mathematics},
  Vol.~\bibinfo{volume}{138}.
\newblock \bibinfo{publisher}{Springer}. xii+534 pages.
\newblock
\showISBNx{3-540-55640-0}
\urldef\tempurl%
\url{https://doi.org/10.1007/978-3-662-02945-9}
\showDOI{\tempurl}


\bibitem[Commelin and Lewis(2021)]%
        {Witt-vectors}
\bibfield{author}{\bibinfo{person}{Johan Commelin} {and}
  \bibinfo{person}{Robert~Y. Lewis}.} \bibinfo{year}{2021}\natexlab{}.
\newblock \showarticletitle{Formalizing the ring of {W}itt vectors}. In
  \bibinfo{booktitle}{\emph{{CPP} '21}} \emph{(\bibinfo{series}{Certified
  Programs and Proofs})}, \bibfield{editor}{\bibinfo{person}{Catalin Hrițcu}
  {and} \bibinfo{person}{Andrei Popescu}} (Eds.). \bibinfo{publisher}{{ACM}},
  \bibinfo{pages}{264--277}.
\newblock
\urldef\tempurl%
\url{https://doi.org/10.1145/3437992.3439919}
\showDOI{\tempurl}


\bibitem[Dedekind(1878)]%
        {Dedekind}
\bibfield{author}{\bibinfo{person}{R. Dedekind}.}
  \bibinfo{year}{1878}\natexlab{}.
\newblock \showarticletitle{Ueber den Zusammenhang zwischen der Theorie der
  ideale und der Theorie der höheren Congruenzen}.
\newblock \bibinfo{journal}{\emph{Abhandlungen der Königlichen Gesellschaft
  der Wissenschaften in Göttingen}}  \bibinfo{volume}{23}
  (\bibinfo{year}{1878}), \bibinfo{pages}{3--38}.
\newblock
\urldef\tempurl%
\url{http://eudml.org/doc/135827}
\showURL{%
\tempurl}


\bibitem[D\'en\`es et~al\mbox{.}(2012)]%
        {maxime}
\bibfield{author}{\bibinfo{person}{Maxime D\'en\`es}, \bibinfo{person}{Anders
  M\"ortberg}, {and} \bibinfo{person}{Vincent Siles}.}
  \bibinfo{year}{2012}\natexlab{}.
\newblock \showarticletitle{A refinement-based approach to computational
  algebra in {C}oq}.
\newblock In \bibinfo{booktitle}{\emph{Interactive theorem proving}}.
  \bibinfo{series}{Lecture Notes in Comput. Sci.}, Vol.~\bibinfo{volume}{7406}.
  \bibinfo{publisher}{Springer, Heidelberg}, \bibinfo{pages}{83--98}.
\newblock
\showISBNx{978-3-642-32347-8; 978-3-642-32346-1}
\urldef\tempurl%
\url{https://doi.org/10.1007/978-3-642-32347-8\_7}
\showDOI{\tempurl}


\bibitem[Dummit and Foote(2004)]%
        {Dummit-and-Foote}
\bibfield{author}{\bibinfo{person}{David~S. Dummit} {and}
  \bibinfo{person}{Richard~M. Foote}.} \bibinfo{year}{2004}\natexlab{}.
\newblock \bibinfo{booktitle}{\emph{Abstract algebra} (\bibinfo{edition}{third
  edition} ed.)}.
\newblock \bibinfo{publisher}{John Wiley \& Sons, Inc.},
  \bibinfo{address}{Hoboken, NJ, USA}. xii+932 pages.
\newblock
\showISBNx{0-471-43334-9}


\bibitem[Dupuis et~al\mbox{.}(2022)]%
        {semilinear}
\bibfield{author}{\bibinfo{person}{Fr\'ed\'eric Dupuis},
  \bibinfo{person}{Robert~Y. Lewis}, {and} \bibinfo{person}{Heather Macbeth}.}
  \bibinfo{year}{2022}\natexlab{}.
\newblock \showarticletitle{Formalized functional analysis with semilinear
  maps}.
\newblock In \bibinfo{booktitle}{\emph{13th {I}nternational {C}onference on
  {I}nteractive {T}heorem {P}roving}}. \bibinfo{series}{LIPIcs. Leibniz Int.
  Proc. Inform.}, Vol.~\bibinfo{volume}{237}. \bibinfo{publisher}{Schloss
  Dagstuhl. Leibniz-Zent. Inform., Wadern}, \bibinfo{pages}{Art. No. 10, 19}.
\newblock
\showISBNx{978-3-95977-252-5}
\urldef\tempurl%
\url{https://doi.org/10.4230/lipics.itp.2022.10}
\showDOI{\tempurl}


\bibitem[Eberl(2020)]%
        {gaussian_integers-isabelle}
\bibfield{author}{\bibinfo{person}{Manuel Eberl}.}
  \bibinfo{year}{2020}\natexlab{}.
\newblock \showarticletitle{Gaussian Integers}.
\newblock \bibinfo{journal}{\emph{Archive of Formal Proofs}}
  (\bibinfo{date}{April} \bibinfo{year}{2020}).
\newblock
\showISSN{2150-914x}
\newblock
\shownote{\url{https://isa-afp.org/entries/Gaussian_Integers.html}, Formal
  proof development}.


\bibitem[Ford(1978)]%
        {Ford}
\bibfield{author}{\bibinfo{person}{David~James Ford}.}
  \bibinfo{year}{1978}\natexlab{}.
\newblock \bibinfo{booktitle}{\emph{O{N} {THE} {COMPUTATION} {OF} {THE}
  {MAXIMAL} {ORDER} {IN} {A} {DEDEKIND} {DOMAIN}}}.
\newblock \bibinfo{publisher}{ProQuest LLC, Ann Arbor, MI}. 145 pages.
\newblock
\urldef\tempurl%
\url{http://gateway.proquest.com/openurl?url_ver=Z39.88-2004&rft_val_fmt=info:ofi/fmt:kev:mtx:dissertation&res_dat=xri:pqdiss&rft_dat=xri:pqdiss:7902127}
\showURL{%
\tempurl}
\newblock
\shownote{Thesis (Ph.D.)--The Ohio State University}.


\bibitem[Futa et~al\mbox{.}(2012)]%
        {gaussian_integers-mizar}
\bibfield{author}{\bibinfo{person}{Yuichi Futa}, \bibinfo{person}{Daichi
  Mizushima}, {and} \bibinfo{person}{Hiroyuki Okazaki}.}
  \bibinfo{year}{2012}\natexlab{}.
\newblock \showarticletitle{Formalization of {Gaussian} integers, {Gaussian}
  rational numbers, and their algebraic structures with {Mizar}}. In
  \bibinfo{booktitle}{\emph{2012 International Symposium on Information Theory
  and its Applications}}. \bibinfo{pages}{591--595}.
\newblock
\showISBNx{978-1-4673-2521-9}


\bibitem[Gr\'egoire and Mahboubi(2005)]%
        {ring}
\bibfield{author}{\bibinfo{person}{Benjamin Gr\'egoire} {and}
  \bibinfo{person}{Assia Mahboubi}.} \bibinfo{year}{2005}\natexlab{}.
\newblock \showarticletitle{Proving equalities in a commutative ring done right
  in {C}oq}.
\newblock In \bibinfo{booktitle}{\emph{Theorem proving in higher order
  logics}}. \bibinfo{series}{Lecture Notes in Comput. Sci.},
  Vol.~\bibinfo{volume}{3603}. \bibinfo{publisher}{Springer, Berlin},
  \bibinfo{pages}{98--113}.
\newblock
\showISBNx{978-3-540-28372-0; 3-540-28372-2}
\urldef\tempurl%
\url{https://doi.org/10.1007/11541868\_7}
\showDOI{\tempurl}


\bibitem[Himmel(2024)]%
        {pratt}
\bibfield{author}{\bibinfo{person}{Markus Himmel}.}
  \bibinfo{year}{2024}\natexlab{}.
\newblock \bibinfo{title}{Formalization of Pratt Certificates}.
\newblock
  \bibinfo{howpublished}{\url{https://github.com/leanprover-community/mathlib4/blob/6439ce3f194a2acd309af6831d753e560c46bcf6/Mathlib/NumberTheory/LucasPrimality.lean}}.
\newblock
\newblock
\shownote{Accessed: 2024-09-06}.


\bibitem[Hogben(2014)]%
        {linalg}
\bibfield{editor}{\bibinfo{person}{Leslie Hogben}} (Ed.).
  \bibinfo{year}{2014}\natexlab{}.
\newblock \bibinfo{booktitle}{\emph{Handbook of linear algebra}
  (\bibinfo{edition}{second} ed.)}.
\newblock \bibinfo{publisher}{CRC Press, Boca Raton, FL}. xxx+1874 pages.
\newblock
\showISBNx{978-1-4665-0728-9}


\bibitem[Lagrange(1798)]%
        {lagrange}
\bibfield{author}{\bibinfo{person}{Joseph-Louis Lagrange}.}
  \bibinfo{year}{1798}\natexlab{}.
\newblock \showarticletitle{Trait{\'e} de la r{\'e}solution des {\'e}quations
  num{\'e}riques, Paris (1798)}.
\newblock \bibinfo{journal}{\emph{{\OE}uvres de Lagrange.}}
  \bibinfo{volume}{8} (\bibinfo{year}{1798}), \bibinfo{pages}{637--649}.
\newblock


\bibitem[Lang(2002)]%
        {Lang}
\bibfield{author}{\bibinfo{person}{Serge Lang}.}
  \bibinfo{year}{2002}\natexlab{}.
\newblock \bibinfo{booktitle}{\emph{Algebra} (\bibinfo{edition}{third edition}
  ed.)}. \bibinfo{series}{Graduate Texts in Mathematics},
  Vol.~\bibinfo{volume}{211}.
\newblock \bibinfo{publisher}{Springer}. xvi+914 pages.
\newblock
\showISBNx{0-387-95385-X}


\bibitem[Lang(2005)]%
        {lang2005undergraduate}
\bibfield{author}{\bibinfo{person}{Serge Lang}.}
  \bibinfo{year}{2005}\natexlab{}.
\newblock \bibinfo{booktitle}{\emph{Undergraduate Algebra}}.
\newblock \bibinfo{publisher}{Spri\-ng\-er}.
\newblock
\showISBNx{9780387220253}
\showLCCN{2004049194}


\bibitem[Lewis(2017)]%
        {Lewis_2017}
\bibfield{author}{\bibinfo{person}{Robert~Y. Lewis}.}
  \bibinfo{year}{2017}\natexlab{}.
\newblock \showarticletitle{An Extensible Ad Hoc Interface between Lean and
  Mathematica}.
\newblock \bibinfo{journal}{\emph{Electronic Proceedings in Theoretical
  Computer Science}}  \bibinfo{volume}{262} (\bibinfo{date}{Dec.}
  \bibinfo{year}{2017}), \bibinfo{pages}{23–37}.
\newblock
\showISSN{2075-2180}
\urldef\tempurl%
\url{https://doi.org/10.4204/eptcs.262.4}
\showDOI{\tempurl}


\bibitem[Lewis and Wu(2022)]%
        {Lewis_bi_directional}
\bibfield{author}{\bibinfo{person}{Robert~Y. Lewis} {and}
  \bibinfo{person}{Minchao Wu}.} \bibinfo{year}{2022}\natexlab{}.
\newblock \showarticletitle{A bi-directional extensible interface between
  {L}ean and {M}athematica}.
\newblock \bibinfo{journal}{\emph{J. Automat. Reason.}} \bibinfo{volume}{66},
  \bibinfo{number}{2} (\bibinfo{year}{2022}), \bibinfo{pages}{215--238}.
\newblock
\showISSN{0168-7433,1573-0670}
\urldef\tempurl%
\url{https://doi.org/10.1007/s10817-021-09611-1}
\showDOI{\tempurl}


\bibitem[{LMFDB Collaboration}(2024)]%
        {lmfdb}
\bibfield{author}{\bibinfo{person}{The {LMFDB Collaboration}}.}
  \bibinfo{year}{2024}\natexlab{}.
\newblock \bibinfo{title}{The {L}-functions and modular forms database}.
\newblock \bibinfo{howpublished}{\url{https://www.lmfdb.org}}.
\newblock
\newblock
\shownote{[Online; accessed 17 September 2024]}.


\bibitem[Mahboubi and Sibut{-}Pinote(2021)]%
        {mahboubi-sibut-pinote-2021}
\bibfield{author}{\bibinfo{person}{Assia Mahboubi} {and}
  \bibinfo{person}{Thomas Sibut{-}Pinote}.} \bibinfo{year}{2021}\natexlab{}.
\newblock \showarticletitle{A Formal Proof of the Irrationality of
  {\(\zeta\)}(3)}.
\newblock \bibinfo{journal}{\emph{Logical Methods in Computer Science}}
  \bibinfo{volume}{17}, \bibinfo{number}{1} (\bibinfo{year}{2021}).
\newblock
\newblock
\shownote{DOI: 10.23638/LMCS-17(1:16)2021}.


\bibitem[Mahboubi and Tassi(2020)]%
        {Assia}
\bibfield{author}{\bibinfo{person}{Assia Mahboubi} {and}
  \bibinfo{person}{Enrico Tassi}.} \bibinfo{year}{2020}\natexlab{}.
\newblock \bibinfo{booktitle}{\emph{Mathematical Components}}.
\newblock \bibinfo{publisher}{Zenodo}.
\newblock
\urldef\tempurl%
\url{https://doi.org/10.5281/zenodo.4282710}
\showDOI{\tempurl}


\bibitem[Neukirch(1999)]%
        {Neukirch}
\bibfield{author}{\bibinfo{person}{J. Neukirch}.}
  \bibinfo{year}{1999}\natexlab{}.
\newblock \bibinfo{booktitle}{\emph{Algebraic number theory}}.
  \bibinfo{series}{Fundamental Principles of Mathematical Sciences},
  Vol.~\bibinfo{volume}{322}.
\newblock \bibinfo{publisher}{Springer}. xviii+571 pages.
\newblock
\showISBNx{3-540-65399-6}
\urldef\tempurl%
\url{https://doi.org/10.1007/978-3-662-03983-0}
\showDOI{\tempurl}
\newblock
\shownote{Translated from the 1992 German original and with a note by Norbert
  Schappacher, With a foreword by G. Harder}.


\bibitem[Pohst and Zassenhaus(1997)]%
        {pohst}
\bibfield{author}{\bibinfo{person}{M. Pohst} {and} \bibinfo{person}{H.
  Zassenhaus}.} \bibinfo{year}{1997}\natexlab{}.
\newblock \bibinfo{booktitle}{\emph{Algorithmic algebraic number theory}}.
  \bibinfo{series}{Encyclopedia of Mathematics and its Applications},
  Vol.~\bibinfo{volume}{30}.
\newblock \bibinfo{publisher}{Cambridge University Press, Cambridge}. xiv+499
  pages.
\newblock
\showISBNx{0-521-59669-6}
\newblock
\shownote{Revised reprint of the 1989 original}.


\bibitem[Rabin(1980)]%
        {Rabin}
\bibfield{author}{\bibinfo{person}{Michael~O. Rabin}.}
  \bibinfo{year}{1980}\natexlab{}.
\newblock \showarticletitle{Probabilistic algorithms in finite fields}.
\newblock \bibinfo{journal}{\emph{SIAM J. Comput.}} \bibinfo{volume}{9},
  \bibinfo{number}{2} (\bibinfo{year}{1980}), \bibinfo{pages}{273--280}.
\newblock
\showISSN{0097-5397}
\urldef\tempurl%
\url{https://doi.org/10.1137/0209024}
\showDOI{\tempurl}


\bibitem[Stevenhagen and Lenstra(1996)]%
        {chebotarev}
\bibfield{author}{\bibinfo{person}{P. Stevenhagen} {and} \bibinfo{person}{H.~W.
  Lenstra, Jr.}} \bibinfo{year}{1996}\natexlab{}.
\newblock \showarticletitle{Chebotar\"ev and his density theorem}.
\newblock \bibinfo{journal}{\emph{Math. Intelligencer}} \bibinfo{volume}{18},
  \bibinfo{number}{2} (\bibinfo{year}{1996}), \bibinfo{pages}{26--37}.
\newblock
\showISSN{0343-6993,1866-7414}
\urldef\tempurl%
\url{https://doi.org/10.1007/BF03027290}
\showDOI{\tempurl}


\bibitem[{The PARI~Group}(2023)]%
        {PARI2}
{The PARI~Group} \bibinfo{year}{2023}\natexlab{}.
\newblock \bibinfo{booktitle}{\emph{{PARI/GP version \texttt{2.15.4}}}}.
\newblock {The PARI~Group}, Univ. Bordeaux.
\newblock
\newblock
\shownote{available from \url{http://pari.math.u-bordeaux.fr/}}.


\bibitem[{The Sage Developers}(2024)]%
        {SageMath}
\bibfield{author}{\bibinfo{person}{{The Sage Developers}}.}
  \bibinfo{year}{2024}\natexlab{}.
\newblock \bibinfo{booktitle}{\emph{{S}ageMath, the {S}age {M}athematics
  {S}oftware {S}ystem ({V}ersion 10.3)}}.
\newblock
\newblock
\shownote{\url{https://www.sagemath.org}. DOI: 10.5281/zenodo.10841614}.


\bibitem[Thiemann et~al\mbox{.}(2020)]%
        {Thiemann-et-al-20}
\bibfield{author}{\bibinfo{person}{Ren{\'e} Thiemann}, \bibinfo{person}{Ralph
  Bottesch}, \bibinfo{person}{Jose Divas{\'o}n}, \bibinfo{person}{Max~W.
  Haslbeck}, \bibinfo{person}{Sebastiaan J.~C. Joosten}, {and}
  \bibinfo{person}{Akihisa Yamada}.} \bibinfo{year}{2020}\natexlab{}.
\newblock \showarticletitle{Formalizing the {LLL} basis reduction algorithm and
  the {LLL} factorization algorithm in {Isabelle}/{HOL}}.
\newblock \bibinfo{journal}{\emph{J. Autom. Reasoning}} \bibinfo{volume}{64},
  \bibinfo{number}{5} (\bibinfo{year}{2020}), \bibinfo{pages}{827--856}.
\newblock
\showISSN{0168-7433}
\urldef\tempurl%
\url{https://doi.org/10.1007/s10817-020-09552-1}
\showDOI{\tempurl}


\bibitem[von~zur Gathen and Gerhard(2013)]%
        {CAS}
\bibfield{author}{\bibinfo{person}{Joachim von~zur Gathen} {and}
  \bibinfo{person}{J\"urgen Gerhard}.} \bibinfo{year}{2013}\natexlab{}.
\newblock \bibinfo{booktitle}{\emph{Modern computer algebra}
  (\bibinfo{edition}{third} ed.)}.
\newblock \bibinfo{publisher}{Cambridge University Press, Cambridge}. xiv+795
  pages.
\newblock
\showISBNx{978-1-107-03903-2}
\urldef\tempurl%
\url{https://doi.org/10.1017/CBO9781139856065}
\showDOI{\tempurl}


\bibitem[Wieser(2021)]%
        {mathlib-scalar-actions}
\bibfield{author}{\bibinfo{person}{Eric Wieser}.}
  \bibinfo{year}{2021}\natexlab{}.
\newblock \showarticletitle{Scalar actions in {Lean}'s mathlib}.
\newblock \bibinfo{journal}{\emph{CoRR}}  \bibinfo{volume}{abs/2108.10700}
  (\bibinfo{year}{2021}).
\newblock
\showeprint[arXiv]{2108.10700}


\end{thebibliography}

\end{document}